\documentclass[%
 aip,
 amsmath,amssymb,
 reprint,%
]{revtex4-1}

\usepackage{graphicx}
\usepackage{dcolumn}
\usepackage{bm}

\usepackage[utf8]{inputenc}
\usepackage[T1]{fontenc}
\usepackage{mathptmx}
\usepackage{etoolbox}

\usepackage[acronym]{glossaries}
\usepackage{xcolor}
\usepackage{booktabs}
\usepackage{multirow}
\usepackage{graphicx}
\usepackage{soul}
\usepackage{bm}
\usepackage{amssymb}
\usepackage{hyperref}

\definecolor{GoodGreen}{rgb}{0,0.6,0}
\definecolor{BadRed}{rgb}{0.8,0.2,0}
\definecolor{purquoise}{rgb}{0.2,0.7,0.65}

\DeclareMathOperator*{\argmin}{\arg\!\min}

\newacronym{ae}{AE}{autoencoder}
\newacronym{cae}{CAE}{convolutional autoencoder}
\newacronym{dl}{DL}{deep learning}
\newacronym{dns}{DNS}{direct numerical simulation}
\newacronym{esn}{ESN}{echo state network}
\newacronym{fr}{FR}{free running}
\newacronym{gru}{GRU}{gated recurrent unit}
\newacronym{relu}{ReLU}{rectified linear unit}
\newacronym{lrelu}{Leaky ReLU}{leaky rectified linear unit}
\newacronym{hae}{H-AE}{hierarchical autoencoder}
\newacronym{lmu}{LMU}{Legendre memory unit}
\newacronym{lsm}{LSM}{liquid state machine}
\newacronym{lstm}{LSTM}{long short-term memory}
\newacronym{mdcnnae}{MD-CNN-AE}{mode decomposing convolutional neural network autoencoder}
\newacronym{mse}{MSE}{mean squared error}
\newacronym{mae}{MAE}{mean absolute error}
\newacronym{nmss}{NMSS}{normalized mean squared similarity}
\newacronym{nn}{NN}{neural network}
\newacronym{nrmse}{NRMSE}{normalized root mean squared error}
\newacronym{pde}{PDE}{partial differential equation}
\newacronym{ode}{ODE}{ordinary differential equation}
\newacronym{pod}{POD}{Proper Orthogonal Decomposition}
\newacronym{pinn}{PINN}{physics-informed neural network}
\newacronym{ra}{Ra}{Rayleigh number}
\newacronym{rl}{RL}{reinforcement learning}
\newacronym{rnn}{RNN}{recurrent neural network}
\newacronym{rbc}{RBC}{Rayleigh-B{\'e}nard convection}
\newacronym{rom}{ROM}{reduced-order model}
\newacronym{tstp}{T-STP}{turbulent superstructure transformation prediction}
\newacronym{sgd}{SGD}{stochastic gradient descent}
\newacronym{smscae}{SMS-CAE}{slim multi-scale convolutional autoencoder}
\newacronym{spiv}{stereo PIV}{stereoscopic particle image velocimetry}
\newacronym{tanh}{tanh}{hyperbolic tangent}
\newacronym{tf}{TF}{teacher forcing}
\newacronym{urnn}{uRNN}{unitary evolution RNN}
\newacronym{vae}{VAE}{variational autoencoder}

\makeatletter
\def\@email#1#2{%
 \endgroup
 \patchcmd{\titleblock@produce}
  {\frontmatter@RRAPformat}
  {\frontmatter@RRAPformat{\produce@RRAP{*#1\href{mailto:#2}{#2}}}\frontmatter@RRAPformat}
  {}{}
}%
\makeatother
\begin{document}

\preprint{AIP/123-QED}

\title[SMS-CAE ROMs]{Slim multi-scale convolutional autoencoder-based reduced-order models for interpretable features of a complex dynamical system}

\author{Philipp Teutsch}%
\affiliation{ 
Group for Data-intensive Systems and Visualization, Technische Universität Ilmenau, D-98684 Ilmenau, Germany 
}%

\author{Philipp Pfeffer}%
\affiliation{ 
Institute of Thermodynamics and Fluid Mechanics, Technische Universität Ilmenau, D-98684 Ilmenau, Germany 
}%

\author{Mohammad Sharifi Ghazijahani}%
\affiliation{ 
Institute of Thermodynamics and Fluid Mechanics, Technische Universität Ilmenau, D-98684 Ilmenau, Germany 
}%

\author{Christian Cierpka}%
\affiliation{ 
Institute of Thermodynamics and Fluid Mechanics, Technische Universität Ilmenau, D-98684 Ilmenau, Germany 
}%

\author{Jörg Schumacher}%
\affiliation{ 
Institute of Thermodynamics and Fluid Mechanics, Technische Universität Ilmenau, D-98684 Ilmenau, Germany 
}%
\affiliation{ 
Tandon School of Engineering, New York University, New York, New York 11201, USA}%

\author{Patrick M{\"a}der}%
\affiliation{ 
Group for Data-intensive Systems and Visualization, Technische Universität Ilmenau, D-98684 Ilmenau, Germany 
}%
\affiliation{ 
Faculty of Biological Sciences, Friedrich-Schiller-Universität Jena, D-07745 Jena, Germany 
}%

\email{philipp.teutsch@tu-ilmenau.de.}

\date{\today}

\begin{abstract}
In recent years, data-driven deep learning models have gained significant interest in the analysis of turbulent dynamical systems. Within the context of reduced-order models (ROMs), convolutional autoencoders (CAEs) pose a universally applicable alternative to conventional approaches. They can learn nonlinear transformations directly from data, without prior knowledge of the system. However, the features generated by such models lack interpretability. Thus, the resulting model is a black-box which effectively reduces the complexity of the system, but does not provide insights into the meaning of the latent features. To address this critical issue, we introduce a novel interpretable CAE approach for high-dimensional fluid flow data that maintains the reconstruction quality of conventional CAEs and allows for feature interpretation. Our method can be easily integrated into any existing CAE architecture with minor modifications of the training process. We compare our approach to Proper Orthogonal Decomposition (POD) and two existing methods for interpretable CAEs. We apply all methods to three different experimental turbulent Rayleigh-B{\'e}nard convection datasets with varying complexity. Our results show that the proposed method is lightweight, easy to train, and achieves relative reconstruction performance improvements of up to $6.4\%$ over POD for $64$ modes. The relative improvement increases to up to $229.8\%$ as the number of modes decreases. Additionally, our method delivers interpretable features similar to those of POD and is significantly less resource-intensive than existing CAE approaches, using less than $2\%$ of the parameters. These approaches either trade interpretability for reconstruction performance or only provide interpretability to a limited extend.
\end{abstract}

\maketitle

\section{\label{sec:introduction}Introduction}
Deep learning methods and algorithms provided numerous new pathways for the analysis and deeper physical insights into the evolution of high-dimensional dynamical systems, such as fluid flows, in the past decade.\citep{brunton2020machine,lu2017reservoir,raissi2020hidden,tian2023lagrangian,fonda2019deep,pandey2020reservoir,pfeffer2022hybrid, teutsch2023data, biferale2023topical,li2024synthetic,salim2024extending} \gls{rbc}, a buoyancy-driven turbulent flow, is one prominent example of such complex dynamical systems. This kind of flow is important for many natural scenarios like for example weather or climate simulations and many engineering applications including heat transfer. Therefore in recent years, related research applied and evaluated the application of deep learning methods like \glspl{cae} as \glspl{rom} for thermal convection flows \citep{vlachas2018data, maulik2021reduced, phillips2021autoencoder, pandey2022direct, yongho2023convolutional}. \gls{cae}-based \gls{rom} are often compared to the data-driven \gls{pod} \citep{berkooz1993proper, weiss2019tutorial, chatterjee2000introduction} approach and have the advantage of providing a nonlinear transformation from the given input space into the reduced lower-dimensional latent space \cite{gonzalez2018deep, ahmed2021nonlinear, obayashi2021feature}. This is attributed to the use of nonlinear activation functions in the corresponding deep neural network architecture. However, in most cases \gls{cae}, similarly to many other deep learning models, operate as black boxes that give few guarantees and make it hard to interpret their inner functioning and latent features. This is an inherent problem of deep learning models, which typically needs to be addressed on a task-specific basis \cite{chakraborty2017interpretability, zhang2018visual, shankaranarayana2019alime}. Further, the \gls{cae} is primarily trained to preserve features that are relevant for an optimal reconstruction regarding its objective function. This, however, may be harmful when the downstream analysis depends on the present small-scale behaviour, such as when modeling the temporal evolution of a nonlinear dynamical system. \gls{pod} enables the interpretation of the reduced space representation by optimal modes w.r.t. the (kinetic) energy. Most of it is contained in some first modes. The primary modes are not necessarily the dynamically most relevant ones, such as for the transition to turbulence in wall-bounded shear flows \cite{eckhardt2007turbulence}.

Alternative solutions have been developed in the past. \citet{ahmed2021nonlinear} proposed a modified nonlinear-\gls{pod} that keeps the first $k$ modes covering $99.9\%$ of the cumulative energy. The authors use the time coefficients of the first $k$ modes as an input representation for a downstream fully-connected \gls{ae}. They compared their approach to a standard \gls{pod} and a conventional \gls{cae}. All three models perform a data reduction to two latent features. For the reconstruction with nonlinear-\gls{pod}, the authors first apply the decoder to receive a prediction of the time coefficients of the first $k$ \gls{pod} modes. Subsequently, they reconstruct the full-order snapshot via reverse \gls{pod}. 

\citet{fresca2022pod} introduced a similar approach to form a POD-deep learning-based reduced order model. However, their model involves the utilization of a \gls{cae} instead of a fully connected \gls{ae} on the time coefficients. Therefore, the time coefficients are reshaped before they are processed by the \gls{cae}. Both approaches take advantage of a linear decomposition to reduce the computational costs for the subsequent deep learning model. As a result, by utilizing these approaches, one can find nonlinear transformations, but will still lose interpretability of the \gls{pod}.

\citet{murata2020nonlinear} proposed a \gls{mdcnnae} as a deep learning approach that has the nonlinear capabilities of a \gls{cae} and allows for latent feature interpretation. Each feature captures information about a separate additive component of the reconstructed flow field snapshot. \gls{mdcnnae} involves the training of a separate sub-decoder for each latent variable. Each sub-decoder produces a snapshot reconstruction component. To retrieve the full order snapshot reconstruction, the components are added together element-wise. \citet{fukami2020convolutional} proposed a variation of \gls{mdcnnae}. Instead of adding the reconstructions of different sub-decoders, they build a \gls{hae} model and training structure. Each hierarchy level has its own latent features but also makes use of the latent features of all previous hierarchy levels. At the same time, each hierarchy level tries to reconstruct the original flow field. This way, the \gls{hae} can learn to combine the latent feature information at different levels in a more complex and nonlinear way. Both approaches were evaluated for a cylinder wake flow at $\text{Re}=100$ and a latent space dimension of $2$ was used.

In this paper, we propose a training strategy for \glspl{cae} that trains the model to extract latent features which are sorted in decreasing order of importance for the reconstruction. While existing CAE-based methods, increase the required resources by orders of magnitude, we will show that our proposed method does not require larger modifications of the network architecture and can be applied to any conventional \gls{cae}. Hence, it only introduces minor modifications in the training process only. During inference, the model can operate just as if it was a conventional \gls{cae}, where it beats the reconstruction performance of existing approaches and \gls{pod} by $4.9\%$ to $229.8\%$. A novel aspect of our approach is that we can decide to further reduce the number of latent features used for a reconstruction later without retraining the autoencoder. In a nutshell, we will preserve a ``modal-type`` information for the features in our modified autoencoder algorithm, which is not available in a standard \gls{cae}.

To give an example, in order to perform time series forecasting of large-scale features of the data as a downstream task, we can simply zero out all other features at smaller scales. For the evaluation of our model, we apply the method to three measured \gls{rbc} datasets at different turbulence intensity which is quantified by the dimensionless Rayleigh number; the data records are taken at $\text{Ra}=9.4\times 10^5$, $2.0\times 10^6$, and $5.5\times 10^6$. In detail, the data records comprise a sequence of $N$ snapshots of the three-dimensional velocity vector fields ${\bm u}(x,y,h/2,t_k)$ with $k=1,\dots N$ at the midplane between the top and bottom plates, at $h/2$. More details will be provided in Sec.~\ref{sec:data}.

The outline of the paper is as follows. We describe the dataset generation in more detail in Sec.~\ref{sec:data}. From there, we continue with a description of the used \gls{cae} architecture and existing approaches in Sec.~\ref{sec:background}. Sec.~\ref{sec:methodology} contains a detailed description of our approach. In Sec.~\ref{sec:evaluation} we evaluate our method on three \gls{rbc} datasets and compare the results to \gls{mdcnnae} and the \gls{hae} approach. We discuss the results in Sec.~\ref{sec:discussion} and end with a conclusion in Sec.~\ref{sec:conclusion}.

\section{Laboratory Measurement Data}\label{sec:data}
The training data, that we use in this paper, stems from a laboratory experiment of a turbulent thermally driven fluid flow. In the following, we describe the flow system and the relevant parameters in detail. We also provide comprehensive information about the data pre-processing and augmentation.

\subsection{Rayleigh-B\'{e}nard convection flow experiment}
\begin{figure}[htb]
    \centering
    \includegraphics[width=\linewidth]{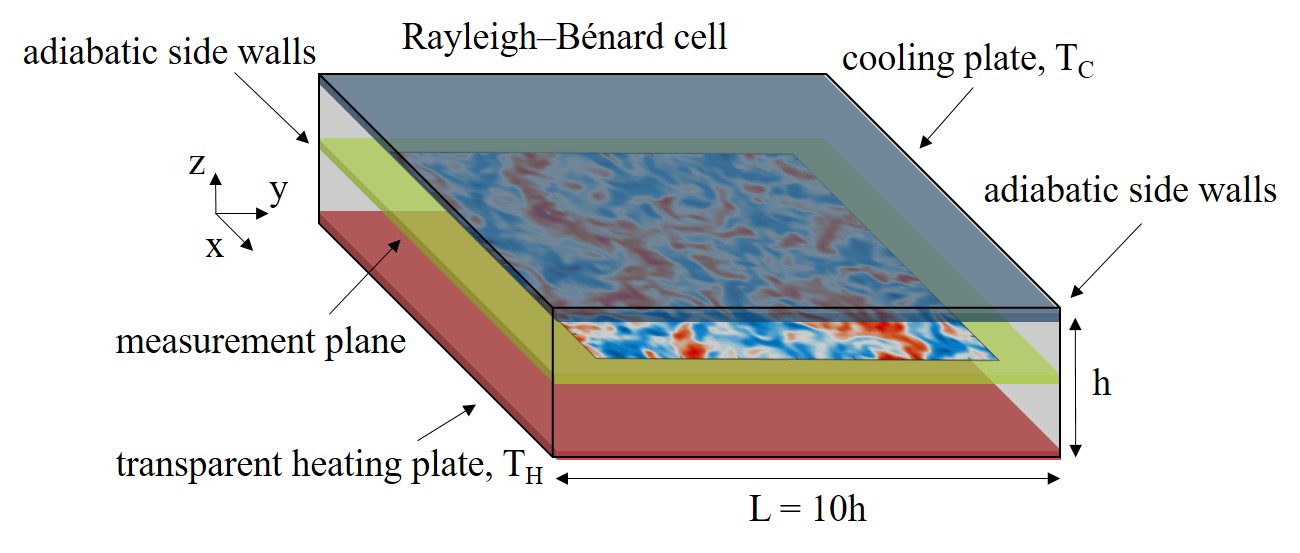}
    \caption{A schematic sketch of the experimental Rayleigh-B\'enard cell. The transparent heating plate provides the optical access to conduct stereoscopic particle image velocimetry in the horizontal mid-plane of the cell. The working fluid, compressed sulfur hexafluoride, is in turbulent motion between the top (in blue) and bottom plates (in red). The side faces, which close the cell, are thermally insulated such that all the supplied heat has to pass through the fluid layer, either by diffusion or by convection.}
    \label{fig:RBCell}
\end{figure}

Turbulent thermal convection is at the heart of various natural flow phenomena, including the Earth's mantle, the outer shell of the Sun, circulations in the oceans, or clouds and cloud clusters in the atmosphere. \gls{rbc} as a turbulent flow between two parallel and horizontally extended solid plates, cf. Fig.~\ref{fig:RBCell} for the flow geometry, is the simplest configuration of this class of flows and thus defines a paradigm~\cite{chilla2012new,verma2018physics}. Turbulent  thermal convection flows, including RBC, form one fundamental class of turbulent flows. The turbulent fluid motion between the plates is driven by the temperature dependence of the mass density in any fluid; lighter hotter fluid rises, heavier colder fluid sinks down. A maintained outer temperature difference between the hot bottom  and colder top plate becomes the permanent driving mechanism of fluid motion (by buoyancy forces) which in addition carries the supplied heat through the layer, from the bottom to the top. For the current study, experimental data of \gls{rbc} has been taken as a representative example of a challenging application for machine learning tasks providing high-dimensional data that follow a complex nonlinear temporal dynamics.

As already stated in Section.~\ref{sec:introduction}, we use three experimental datasets of convection in pressurized sulfur-hexafluoride (SF\textsubscript{6}), for Rayleigh numbers of $\text{Ra} = 9.4\times 10^5$, $2.0\times 10^6$ and $5.5\times 10^6$. The Rayleigh number is a dimensionless parameter which relates the buoyancy and viscous forces of the fluid. As Ra increases, the complexity of the turbulent flow increases as well. It is given by $\text{Ra}=g\alpha \Delta T h^3/(\nu\kappa)$. Here, $\Delta T=T_{\rm bot}-T_{\rm top}>0$ is the temperature difference between the uniformly heated and cooled plates. The other quantities in the definition of Ra are the fluid properties namely, thermal expansion coefficient $\alpha$, kinematic viscosity $\nu$, and thermal diffusivity $\kappa$ together with the acceleration due to gravity, $g$. Two more parameters characterize the convective flow. The Prandtl number $\text{Pr}=\nu/\kappa=0.7$ and the aspect ratio of the cell, which related lateral extension to height and is in our case $\Gamma = L/h = 10$. 

\begin{figure*}[htb]
    \centering
    \includegraphics[width=0.7\linewidth]{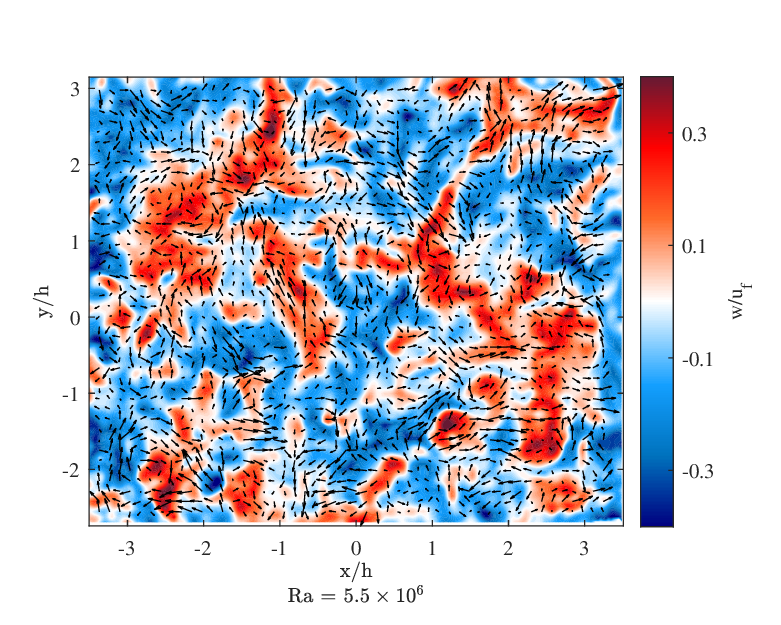}
    \caption{Instantaneous turbulent velocity field  in the horizontal mid-plane for a Rayleigh number Ra $= 5.5\times 10^6$. The color map indicates the out-of-plane velocity component $w$, while the in-plane components $u$ and $v$ are represented by the vectors. Note that only one out of four vectors in each direction is plotted.}
    \label{fig:uvw}
\end{figure*}

We conducted \gls{spiv} in the horizontal mid-plane in order to measure the in-plane velocity components ($u$ and $v$) along with the out-of-plane velocity component $w$. Each of the 3 datasets has $20\,000$ snapshots with $10$ Hz recording frequency and the field of view which covers an area of $A = 7h \times 5.8h$ at the center of the cell, see again Fig. \ref{fig:RBCell}. There are $174 \times 208$ velocity vectors that define the flow field with a spatial resolution of $1 \times 1$ mm$^2$. Figure \ref{fig:uvw} shows an example of a measured instantaneous velocity field. Here, the vectors indicate the in-plane velocity components $u$ and $v$, while the background color map shows the out-of-plane velocity component $w$. The values are normalized to the free-fall velocity $u_f=\sqrt{\alpha\Delta Tgh}$. In order to further visualize the turbulent flow and to illustrate its chaotic behavior, a video of the flow field dynamics is available in the supplemental online material. Further details about the measurement setup and the physics of the flow can be found in \citet{sharifi2023scalex} and \citet{sharifi2024spatio}, respectively.

For the experimental evaluation of our model, we will use the out-of-place velocity component $w$ only. We apply some data pre-processing and use a version with a reduced spatial resolution of $87\times104$ pixels. This is done to save memory and computation time. Finally, we trim the range of data values ($w/u_f$) between $-0.6$ and $0.6$ to remove outliers. We set them to $-0.6$ and $0.6$ respectively. The trimming affects $0.007\%$, $0.001\%$ or $0.005\%$ of all values, depending on the dataset. During training and testing, the input snapshots are generated by applying a $64\times64$ pixel crop on the reduced snapshots. This allows us to use data augmentation for the deep learning models.

\subsection{Augmentation of the experimental snapshot data}
\begin{figure}[htb]
    \centering
    \includegraphics[width=0.85\linewidth]{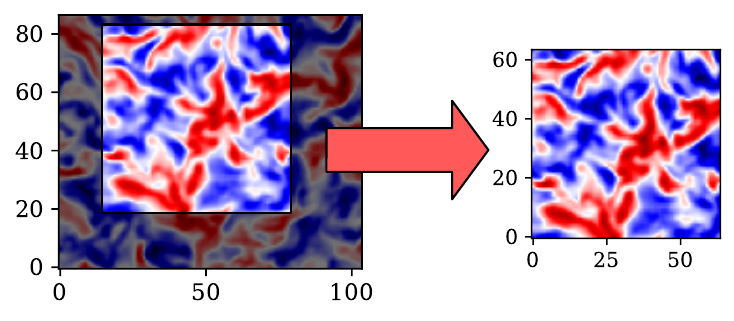}
    \caption{Sketch of the shift and crop operation applied to an example snapshot for data augmentation.}
    \label{fig:data_aug}
\end{figure}

When training deep learning models, we use data augmentation as a regularization measure. Although it is not mandatory to apply data augmentation during training, previous studies have shown, that it improves the generality of the resulting deep learning model and helps to avoid overfitting. \cite{perez2017effectiveness,cubuk2019autoaugment,yang2022image,teutsch2024large}
As augmentation operations, we use a shift operation in horizontal and vertical direction followed by a crop operation, described above (cf. Sec.~\ref{sec:data}). We take advantage of the fact that the original dimensions of our data snapshots are larger than the input dimensions of our models. Data augmentation is only applied during training phase; it is omitted for validation and testing phase. Here, we always perform a center crop to match the desired snapshot dimensions. Figure~\ref{fig:data_aug} shows an example snapshot before and after the shift and crop operations. The data augmentation operations are applied for each training snapshot separately. The shift width in either direction is chosen at random each time.

\begin{figure*}[htb]
    \centering
    \includegraphics[width=0.8\linewidth]{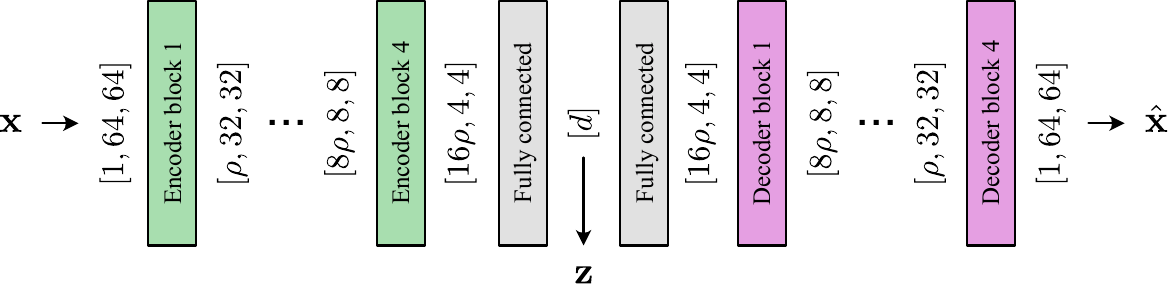}
    \caption{Basic convolutional autoencoder (CAE) architecture. Illustrated are the data dimensions at the different stages of the encoding and decoding process. The CAE takes a snapshot $\mathbf{x}$ and encodes it into a latent representation $\mathbf{z}$. Subsequently it is expanded into the reconstructed snapshot $\hat{\mathbf{x}}$. The squared brackets contain the shape of input, output and features on the forward pass. Here, $\rho$ is a hyper-parameter that controls the number of filters in the convolutional layers. Thus it determines the number of channels of the corresponding output feature maps.}
    \label{fig:cae_architecture}
\end{figure*}

\section{\label{sec:background}Existing Data-driven reduction methods}
In this section, we give a description of the existing data-driven methods which are used in our evaluation. We introduce the \gls{pod} that acts as a baseline and the \gls{cae} architecture that we use as the foundation for different nonlinear data-driven \gls{rom} approaches that will be specified in subsection C and D.

\subsection{Proper Orthogonal Decomposition by snapshot method}
The Proper Orthogonal Decomposition\cite{LIANG_POD} (POD) is a linear data decomposition method which transforms a data matrix $M \in \mathbb{R}^{T,\mathcal{X}}$ that contains $T$ measurements (or snapshots) of $\mathcal{X}$ spatial features (e.g, pixels, grid points or resolved flow components) to temporal modes and spatial modes. This can be interpreted as a singular value decomposition of the data matrix which is given by
\begin{equation}
M = U_T \ \Sigma \ U_\mathcal{X}\,,
\end{equation}
where $U_T \in \mathbb{R}^{T,T}$ is an orthogonal matrix storing $T$ temporal modes for $T$ time steps, $\Sigma \in \mathbb{R}^{T,\mathcal{X}}$ is a diagonal matrix storing the descending sorted singular values or principal components of the data and $U_\mathcal{X} \in \mathbb{R}^{\mathcal{X},\mathcal{X}}$ is an orthogonal matrix storing $\mathcal{X}$ spatial modes with $\mathcal{X}$ spatial components. A spatial mode corresponds herein to a static image of the out-of-plane velocity component. The first spatial mode is scaled with the first and thereby largest singular value, which has the largest contribution to the total kinetic energy. The corresponding temporal mode indicates how strongly this spatial mode is present in each of the $T$ measurements.

In this work, $\mathcal{X} \ll T$, which implies  that the rectangular and diagonal matrix $\Sigma$ only stores $\mathcal{X}$ singular values. Thereby, we can reduce $\Sigma$ to a square matrix of size $\mathcal{X}$ and $U_T$ to a rectangular matrix in $\mathbb{R}^{\mathcal{X},T}$ to reconstruct $M$. We use $T=16\,000$ measurements of $\mathcal{X}= 4\,096$ spatial features (or grid points), leading to $4\,096$ temporal and spatial modes. We identify the spatial modes $U_\mathcal{X}$ and principal components $\Sigma$ by the covariance matrix of the mean-centered data matrix $\bar{M}$ by
\begin{equation}
C = \frac{1}{\mathcal{X}-1} \ \bar{M}^T\bar{M} = \frac{1}{\mathcal{X}-1} \  U_\mathcal{X}^T \Sigma^2 U_\mathcal{X}\,.
\end{equation}
For the temporal modes, we use $U_T= \bar{M}U_\mathcal{X}^T \Sigma^{-1}$. 

\subsection{Convolutional autoencoder}
We compare different \gls{cae} approaches to the conventional \gls{pod} analysis. For these, we use a conventional \gls{cae} with a customized architecture as base for all other \gls{cae}-based approaches. An autoencoder network in general consists of two parts, the encoder and the decoder part. The encoder part performs a function $E_{d, \theta}: \mathbb{R}^{h, w} \rightarrow \mathbb{R}^{d}$ that takes a data snapshot $\mathbf{x}$ as parameter and transforms it into a latent representation $\mathbf{z}$, where $h$ and $w$ are the vertical and horizontal dimension of the input snapshot. The decoder performs a function $D_{d, \theta}: \mathbb{R}^{d} \rightarrow \mathbb{R}^{h,w}$. It takes a latent representation $\mathbf{z}$ and transforms it into the reconstruction of the data snapshot $\mathbf{\hat{x}}$. Hence, the autoencoder can be described by a function $F_{d, \theta}: \mathbb{R}^{h,w} \rightarrow \mathbb{R}^{h, w}$ that is composed of $E_{d, \theta}$ and $D_{d, \theta}$. Each function represents a neural (sub-)network that depends on a parameter set $\theta$. We omit the $\theta$ subscript in the rest of this paper for the sake of simplicity. Thus, we write $F_d$, $E_d$ and $D_d$ and only add $\theta$ when needed. As a result, the function of the autoencoder can be written as follows:
\begin{equation}
    F_{d}(\mathbf{x}) = (E_{d} \circ D_{d})(\mathbf{x}) = \hat{\mathbf{x}}\,.
\end{equation}
The encoder $E_d$, used in this paper, consists of four encoder blocks, each of which utilizes two 2d-convolutional layers with \gls{lrelu} \cite{maas2013rectifier} activation and a subsequent batch normalization \cite{ioffe2015batch} layer. The convolutional layers have kernel size $3\times3$. Each encoder block ends with a max-pooling layer with kernel size $2\times2$ to perform the data reduction and a dropout layer as regularization measure. After the last encoder block, we perform a flatten operation that transforms the 3-dimensional output feature map into a 1-dimensional feature vector. The vector is processed through two fully connected layers with another batch normalization layer in between. These layers reduce the feature vector dimension to finally fit the desired latent dimension $d$. The last fully connected layer is followed by a \gls{tanh} activation.

The decoder $D_d$ is constructed analogously to $E_d$ in reverse order. It starts with two fully connected layers with \gls{lrelu} activation and batch normalization. Their output is brought into the desired 3-dimensional shape to serve as input for the first of four decoder blocks. The first three decoder blocks consist of a nearest neighbor up-sampling layer with a kernel size $2\times2$ to perform the data expansion. It is followed by two 2d-convolutional layers with \gls{lrelu} activation and batch normalization. Here the kernel size is also $3\times3$. The blocks end with another dropout layer. The last decoder block is built differently. Its second convolutional layer directly outputs the reconstructed data snapshot through a sigmoid activation function. Figure~\ref{fig:cae_architecture} provides an overview of the \gls{cae} structure and its dimensionality. The hyper-parameter $\rho$ serves as scale factor for the number of channels of the feature maps between the different blocks.

The architectural and hyper-parameter choices made in this paper are not generalizable. They depend on the specific learning task, the available computational resources, and the complexity and size of the data. Therefore, different choices may be advised in other scenarios.

A \gls{cae} is usually trained in a self-supervised fashion, where the network's input is also its target. Thus, it aims to optimize the following problem:
\begin{equation}
    \argmin_{\theta} \mathcal{L}(D_d(E_d(\mathbf{x})), \mathbf{x})\,,
\end{equation}
where $\mathcal{L}$ is the objective function to reduce. The training is considered successful when $\mathbf{x} \approx \hat{\mathbf{x}}$.

\begin{figure*}[htb]
    \centering
    \includegraphics[width=0.85\linewidth]{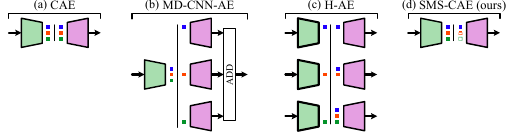}
    \caption{Minimal schematic architectures that we study in this paper. Each model is visualized with only three latent features for illustration. Per architecture, the left hand side represents the features as they leave the encoder and while the right hand side shows how they enter the decoder. (a) is the basic CAE that is also shown in Figure~\ref{fig:cae_architecture} in detail. It simply forwards the latent features as they are. (b) and (c) show the existing interpretable CAE methods that have sub-encoders or -decoders that are responsible for a specific subset of features. (c) represents our method that randomly sets values to zero at the lower end of the latent feature vector during training.}
    \label{fig:cae_mini_archs}
\end{figure*}

In this and the following Section, we will introduce different \gls{cae} approaches that use a conventional \gls{cae} as foundation. We present a simplified sketch for each of them, to provide an intuition for their architecture, cf. Fig.~\ref{fig:cae_mini_archs}.

\subsection{Mode decomposing convolutional neural network autoencoder}
The mode decomposing convolutional neural network autoencoder (MD-CNN-AE) as proposed by \citet{murata2020nonlinear}, is an approach for a \gls{cae}-based nonlinear \gls{rom} that introduces an adapted architecture. While the encoder $E_d$ remains unchanged in comparison to the conventional \gls{cae}, the decoder $D_d$ is split into multiple sub-decoders $D_{i,d}$, where the number of decoders corresponds to the latent data space dimension $d$. Each sub-decoder learns to reconstruct a different additive component of the target snapshot, based on a separate latent feature $z_i\in \mathbf{z}$. Hence, we can formulate the decoder as
\begin{equation}
    D_d(\mathbf{z}) = \sum_{i=1}^{d} D_{i,d}(z_i)\,.
\end{equation}
Given $E_d$ and $D_d$, the optimization problem remains similar to the base \gls{cae}.

\subsection{Hierarchical autoencoder}
The hierarchical autoencoder (H-AE) approach by \citet{fukami2020convolutional} implies to train a set of sub-encoders $\tilde{E}_{i, c_i}$ and sub-decoders $\tilde{D}_{i, d_i}$ in a hierarchical scheme to achieve an interpretability of the latent features. Sub-encoder $\tilde{E}_{i, c_i}$ produces $c_i$ output features and sub-decoder $\tilde{D}_{i, d_i}$ takes $d_i$ latent features as input. In the first stage, the training starts with sub-encoder $\tilde{E}_{1, c_1}$ and sub-decoder $\tilde{D}_{1, d_1}$, where $c_1 = d_1$. As soon as the training converges, their weights are freezed and remain fixed for the rest of the training. With each following stage $i$, another pair of sub-encoder and decoder $\tilde{E}_{i, c_i}$, $\tilde{D}_{i, d_i}$ is trained. Sub-decoder $\tilde{D}_{i, d_i}$ receives $c_i$ latent features from $\tilde{E}_{i, c_i}$ together with those of all previously trained sub-encoders, thus $d_i = \sum_{j=1}^{i}c_i$. The authors used $c_i = 1$ in their paper. Thus, $d$ is not only the latent size, but also the number of sub-autoencoders. Therefore, we reduce the symbols for sub-encoder and sub-decoder to $\tilde{E}_{i}$ and $\tilde{D}_{i}$ for simplicity. Finally, the trained model can be written as:
\begin{equation}
    F_{d}(\mathbf{x}) = \tilde{D}_{d}([\tilde{E}_{i}(\mathbf{x}); \dots; \tilde{E}_{d}(\mathbf{x})])\,.
\end{equation}
During inference, we only require the last sub-decoder, but all sub-encoders. Due to the hierarchical training structure, we can also neglect some of the later latent features and still receive a valid reconstruction. For example, given that $d>2$ and we want to use only $2$ latent features, we use the second decoder and the first two encoders:
\begin{equation}
    F_{2}(\mathbf{x}) = \tilde{D}_{2}([\tilde{E}_{1}(\mathbf{x}); \tilde{E}_{2}(\mathbf{x})])\,.
\end{equation}
The minimization problem addressed differs from that of the base \gls{cae}. In hierarchical training, it is instead approached as a series of minimization problems. Except for the first stage, the minimization problem at each successive stage $i$ can be formulated as
\begin{equation}
    \argmin_{\theta_i} \mathcal{L}(\tilde{D}_{i}([\tilde{E}_{1}(\mathbf{x}); \dots; \tilde{E}_{i-1}(\mathbf{x}); \tilde{E}_{i}(\mathbf{x})]), \mathbf{x})\,,
\end{equation}
where $\theta_i$ represents the parameter set of the sub-encoder and decoder for stage $i$. The solution must also take the earlier sub-encoders $\tilde{E}_{1}, \dots, \tilde{E}_{i-1}$ into account, which are dependent on the parameter sets $\theta_1, \dots, \theta_{i-1}$. Therefore, $\theta_1, \theta_2, \dots$ are optimized sequentially, with each subsequent stage utilizing the optimal parameter sets found for the previous stages.

\section{\label{sec:methodology} Slim multi-scale convolutional autoencoder}
We propose a novel approach which creates a \gls{smscae}. \gls{smscae} is inspired by \gls{pod} and the interpretability of its modes. \gls{pod} modes with higher kinetic energy hold information about the structures at the larger scale of the full system, while the energy proportion decreases with increasing mode index (since they are ordered with respect to their kinetic energy). Analogously, we propose to train a \gls{cae} that captures as much information as possible in the first latent features. In contrast to \gls{hae}, which comes with a heavy increase of required training resources, \gls{smscae} is lightweight in memory and time resources. While \gls{hae} uses a number of sub-models that scale linearly with the number of latent features, \gls{smscae} requires only a minimal architectural adaption to the base model. The training of the \gls{hae} is performed in a hierarchical, stage-wise fashion, where sub-models are trained in series which results in a longer training time. The training of the \gls{smscae} optimizes all model parameters in each gradient descent step just like the original base model. This avoids the imbalance between the reconstruction quality of the large-scale and small-scale spatial features, that may appear due to \gls{hae} training. The core working principle of \gls{smscae} is inspired by the well-known dropout regularization \cite{srivastava2014dropout}. Dropout helps to provide an effect comparable with ensembles training without the overhead of maintaining multiple models \cite{bachman2014learning, hara2016analysis}. Similarly, we employ a restricted dropout mechanism that helps our model to train multiple sub-tasks at the same time and avoids the overhead associated with a hierarchical model. In contrast to an hierarchical training scheme, we allow the model to address all sub-tasks equally, thus preventing an imbalance in reconstruction quality. The constrained dropout mechanism is positioned immediately after the last encoder layer and sets a random number of latent features to zero during training. Hence, on the forward pass, we preserve a contiguous subset of features that always includes the first one. Thus, if $\mathbf{z} = (z_1, z_2, \dots, z_d)$ is the latent feature vector, the decoder uses features with lower indices more often than those with higher indices. As a result, the model learns to encode the most important information in the first, more frequently trained features.

As described in Section~\ref{sec:background}, $F_d$ represents the function of a conventional autoencoder with latent dimension $d$. For our proposed \gls{smscae}, we introduce a restricted dropout function $\varphi_{\hat{d}}$:
\begin{equation}
    \varphi_{\hat{d}}(\mathbf{z}) = \mathbf{z} \odot [\mathbf{1}_{\hat{d}};\mathbf{0}_{d-\hat{d}}]\,,
\end{equation}
where latent vector $\mathbf{z} \in \mathbb{R}^d$, $\hat{d} \leq d$, $\mathbf{1}_{\hat{d}} \in \{1\}^{\hat{d}}$ and $\mathbf{0}_{d-\hat{d}} \in \{0\}^{d-\hat{d}}$. Thus, $\varphi_{\hat{d}}$ basically preserves the first $\hat{d}$ features of $\mathbf{z}$ and turns the remaining into $0$. The constrained dropout mechanism $\varphi_{\hat{d}}$ is placed between the encoder $E_d$ and decoder $D_d$ of the \gls{cae} to restrict the information available for reconstruction. Thus, the \gls{smscae} is formalized as
\begin{equation}
    F_{d,\hat{d}}(\mathbf{x}) = (E_d \circ \varphi_{\hat{d}} \circ D_d)(\mathbf{x})\,.
\end{equation}
During training, $\hat{d}$ is a random variable drawn from a discrete uniform distribution on $[1, d]$. Hence, for $1/d$ of all samples, the model processes throughout training, it tries to encode into the first latent feature $z_1$ only. Therefore, $z_1$ holds as much information as possible on the large-scale structures of the snapshots. Additionally, the model tries to incorporate information of $z_1$ into the reconstruction of small-scale structures in the cases where more latent features are available. As a result, the model addresses the following optimization problem
\begin{equation}
    \argmin_{\theta}(\sum_{\hat{d}=1}^d \mathcal{L}(F_{d, \hat{d}}(\mathbf{x}), \mathbf{x})).
\end{equation}
\begin{figure}[htb]
    \centering
    \includegraphics[width=0.85\linewidth]{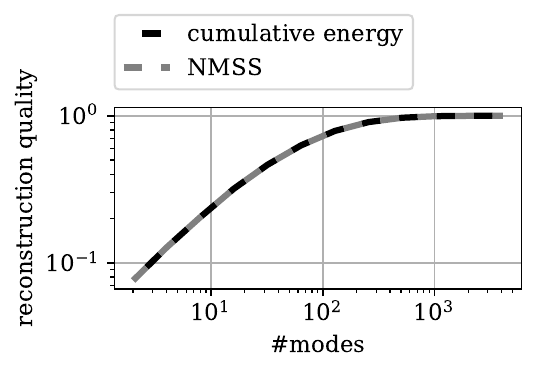}
    \caption{Reconstruction quality for training data for the experiment at a Rayleigh number of $\text{Ra}=9.4\times 10^5$ displayed as cumulative energy and normalized mean squared similarity (NMSS).}
    \label{fig:nmss_cumul_energy}
\end{figure}
The \gls{mse} is a common choice of objective function $\mathcal{L}$ for \glspl{cae}, and is considered reasonable for pure reconstruction tasks. However, \gls{smscae} provides interpretability in terms of importance for reconstruction according to the objective function. Thus, the objective function is a hyper-parameter that mainly controls the desired interpretability. For example, if our goal is to encode information about the flow at the boundaries, we can alter the objective function to give more weight to the reconstruction error at the edges. The model will then be trained to encode boundary region information with higher priority in the first feature(s). It is also possible to adjust the objective depending on how many features are passed for a given sample to give certain features a very specific additional task. This shows that there are massive possibilities to tune the interpretability of the \gls{rom} dimensions beyond minimizing the squared reconstruction error. However, this analysis is beyond the focus of this paper and should be addressed in future research.

\section{\label{sec:evaluation}Evaluation}
In the evaluation, we compare \gls{pod} and the three \gls{cae}-based \gls{rom} approaches discussed above. To ensure the comparability of the reconstruction results, we take specific measures addressing the data split and the evaluation metric.

\subsection{Experimental setup}
For deep learning models, the training is typically structured into three phases: training, validation and testing, where each phase uses a completely separate part of the whole dataset. The model solely adapts its weights on the training set, while the validation dataset can be used to compute a validation loss to decide whether the model is done training or to adjust hyper-parameters. To compare the model with others, the model is applied on the test set and the performance metric is calculated. To keep the \gls{cae} results comparable to \gls{pod} we will follow this procedure for all methods. We apply an $80-10-10$ split to divide each dataset into a training, validation and test set. Thus, also the \gls{pod} modes and singular-values are retrieved solely relying on the information from the training set. 

As a second measure, we aim to establish an evaluation metric to compare the reconstructions of the \gls{pod} with those of the deep learning methods. In the case of the \gls{pod}, we can use the retrieved singular-values to judge the amount of cumulative energy information preserved after discarding a certain number of modes. For \glspl{cae} based methods, we do not have this information. Instead, an error metric based on \gls{mse} or \gls{mae} is calculated. It is common practice to normalize the error with the mean or variance of the data. We use the \gls{mse} normalized by the average variance per grid point. Finally, we subtract the resulting error from $1$. Thus, a large score refers to a high similarity. We refer to the resulting similarity metric as \gls{nmss}. The \gls{nmss} behaves almost similar to the the preserved cumulative energy curve of the \gls{pod} above $0$:
\begin{equation}
    \text{NMSS}(\mathbf{x}, \hat{\mathbf{x}}) = 1 - \frac{\sum_{i = 1}^{N_y}\sum_{j = 1}^{N_x} (\mathbf{x}_{i, j} - \hat{\mathbf{x}}_{i, j})^2}{\sum_{k=1}^{N_y}\sum_{l=1}^{N_x}(\bar{\mathbf{x}}_{k, l} - \mathbf{x}_{k, l})^2}\,,
\end{equation}  
where $\bar{\mathbf{x}}_{k, l}$ is the average value of the training snapshots at position $(k, l)$. The \gls{nmss} ranges from $-\infty$ to $1.0$, where values smaller than $0.0$ imply a result that is considered worse than using the average training snapshot as reconstruction. An \gls{nmss} value of $1.0$ stands for a perfect reconstruction. Figure~\ref{fig:nmss_cumul_energy} shows the \gls{nmss} of the \gls{pod} reconstructions on different numbers of modes on the training data at $\text{Ra}=9.4\times 10^5$. It also plots the cumulative energy based on the corresponding singular values. The \gls{nmss} curve almost exactly follows the cumulative energy curve with only a small deviation that is negligible. Therefore, we use \gls{nmss} as an evaluation metric for \gls{cae} and \gls{pod} reconstruction.

For the batch-wise training, we used the Adam optimizer \cite{kingma2014adam} with a learning rate of $10^{-5}$ and a batch size of $128$ snapshots. Additionally, we applied a learning rate scheduler to adapt the learning rate in the later stages of training. The learning rate scheduler reduces the learning rate to $10\%$ when the validation loss is stagnant for $30$ epochs. To calculate the training and validation loss, we used the \gls{mse} as loss function. The hyper-parameter $\rho$ which controls the number of filters in the convolutional layers (cf. Fig.~\ref{fig:cae_architecture}) was set to $32$. All hyper-parameters were selected during a greedy search and are fixed for all \gls{cae}-based models in all experiments. Before we feed the snapshots to a model, we normalized the dataset between $0.0$ and $1.0$. To calculate the evaluation metric, we reverse the normalization of the reconstructions and compare them to unnormalized target snapshots.

\subsection{Baseline experiments: POD vs. CAE}
\begin{figure*}[htb]
    \centering
    \includegraphics[width=0.95\linewidth]{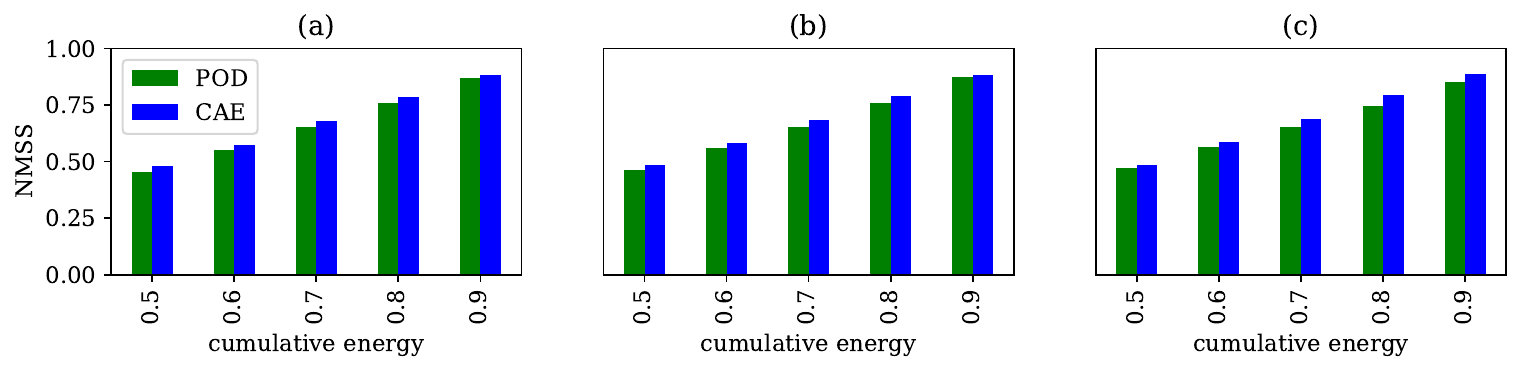}
    \caption{Average test normalized mean squared similarity (NMSS) of POD and CAE reconstructions on a fixed set of energy preservation levels for each dataset: (a) $\text{Ra}=9.4\times 10^5$, (b) $\text{Ra}=2.0\times 10^6$, (c) $\text{Ra}=5.5\times 10^6$.}
    \label{fig:nmss_bars}
\end{figure*}

In our first set of experiments, we compare the reconstruction performance of the conventional \gls{cae} and \gls{pod} for different latent sizes and respective numbers of modes. We choose them according to the preserved cumulative energy to be at levels $0.5$, $0.6$, $0.7$, $0.8$ and $0.9$. We list the corresponding numbers of modes for each dataset in Table~\ref{tab:modes_for_energy}. We refer to these as energy levels. Figure~\ref{fig:nmss_bars} shows the average test \gls{nmss} at different energy levels for each dataset. It becomes clear that the \gls{cae} consistently surpasses the \gls{pod} reconstruction performance throughout all energy levels and datasets. However, it is apparent that the absolute benefit of the \gls{cae} shrinks towards higher energy levels. We set the number of latent features according to the number of \gls{pod} modes required to preserve the same amount of energy for each dataset. Nevertheless, the reconstruction quality on the test snapshots shrinks with increased $\text{Ra}$, which is due to the higher complexity of the more turbulent flow at higher Ra. Consistently, we observe this effect for both approaches. 

\begin{table}[htb]
    \centering
    \begin{tabular}{r r r r}
    \toprule
         Cumulative      & \multicolumn{3}{c}{Ra}      \\
         energy          & $9.4\times 10^5$ & $2.0\times 10^6$ & $5.5\times 10^6$ \\
    \midrule
         $0.5$            &  $38$   &  $37$  &    $35$  \\
         $0.6$            &  $57$   &  $58$  &    $57$  \\
         $0.7$            &  $86$   &  $91$  &    $95$  \\ 
         $0.8$            & $136$   &  $149$ &   $166$  \\
         $0.9$            & $245$   &  $280$ &   $333$  \\
    \bottomrule
    \end{tabular}
    \caption{Number of modes required for each dataset to reconstruct a certain proportion of the cumulative energy of the full-order model}
    \label{tab:modes_for_energy}
\end{table}

\begin{figure*}[htb]
    \centering
    \includegraphics[width=0.75\linewidth]{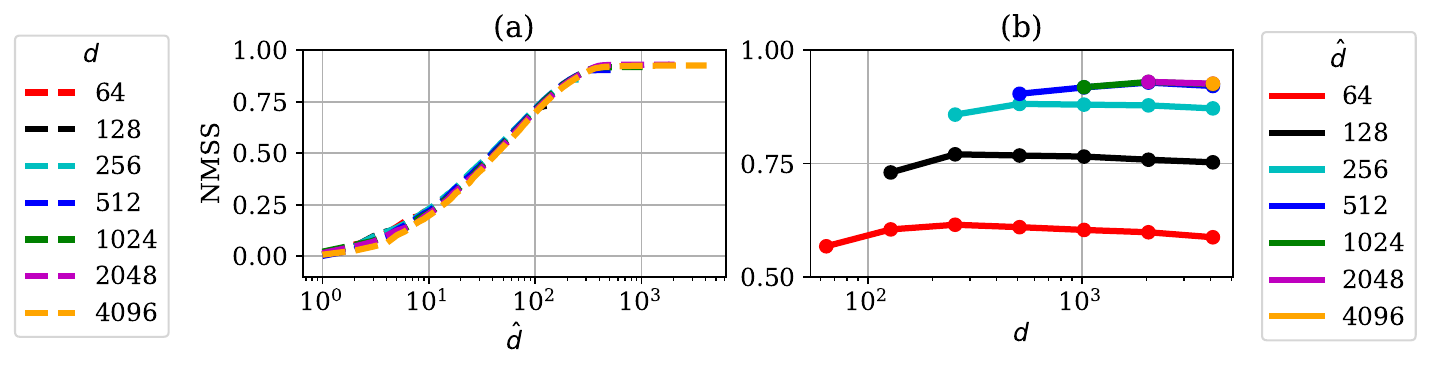}
    \caption{Test NMSS of the SMS-CAE with different amounts of latent features in the training, $d$, and test phases, $\hat{d}$, for a Rayleigh number of $\text{Ra}=9.4\times 10^5$.}
    \label{fig:smscae_nmsss}
\end{figure*}

\subsection{Characteristics of SMS-CAE}
For a conventional \gls{cae} the latent dimension is set during training and remains fixed. All latent features contribute approximately equally to the reconstruction and we cannot easily create a minimal sub-selection of features that cover most information. \gls{smscae} approaches this issue with the separation of latent feature sub-sets during training. As a result, after a training with $d$ latent features, we can use the same model to produce reconstructions from $\hat{d} = 1, \dots, d$ features without retraining. More specifically, we assume that an \gls{smscae} model trained on $d$ latent features is as good at reconstructing from $\hat{d} = \lfloor d/2\rfloor$ features as an \gls{smscae} model that was originally trained on that many features. To test whether this assumption holds, we perform a second set of experiments. We train the \gls{smscae} on $d = 64, 128, 256, 512, 1024, 2048, 4096$ latent features. Subsequently, we apply the model on the test set to reconstruct snapshots from $\hat{d} = 1, \dots, d$ features respectively. The results reveal that the approximate progression of NMSS is similar regardless the number of latent features in training, cf. Fig.~\ref{fig:smscae_nmsss}(a). Furthermore, we see that for $\hat{d} \gtrsim 400$ features the improvement saturates at an \gls{nmss} of $\approx0.93$. In Fig.~\ref{fig:smscae_nmsss}(b), we see how well the model can reconstruct from $\hat{d}$ features depending on what $d$ it was trained before. The surprising observation is that the model is not best when $\hat{d}=d$. In the tested cases, it seems to be advised to train the model on a $d$ that is $2$ to $4$ times larger than the $\hat{d}$ used during inference.

\begin{table*}[htb]
    \centering
    \begin{tabular}{r l r | c c | c c | c c}
    \toprule
                &            & Parameters  & \multicolumn{2}{c|}{$\text{Ra}=9.4e5$}  & \multicolumn{2}{c|}{$\text{Ra}=2.0e6$} & \multicolumn{2}{c}{$\text{Ra}=5.5e6$} \\
                &            &             & NMSS    & Rel. impr.                 & NMSS        &  Rel. impr.           & NMSS        &  Rel. impr.              \\
    \midrule
                & POD        & ($266$K)    & $0.578$ & --                         & $0.582$     & --                    & $0.583$     & --                            \\
                & CAE        & $19$M       & $0.608$ & $+5.3\%$                   & $0.600$     & $+3.1\%$              & $0.611$     &  $+5.0\%$                    \\
                & MD-CNN-AE  & $586$M      & $0.610$ & $+5.5\%$                   & $0.616$     & $+5.8\%$              & $0.610$     &  $+4.8\%$                    \\
                & H-AE       & $1.2$B      & $0.489$ & $-15.3\%$                  & $0.530$     & $-9.0\%$              & $0.534$     &  $-8.2\%$                     \\
    \midrule
         (ours) & SMS-CAE    & $21$M       & $0.615$ & $+6.4\%$                   & $0.611$     & $+4.9\%$              & $0.617$     &  $+5.9\%$                      \\
    \bottomrule
    \end{tabular}
    \caption{NMSS and relative improvement on the test data reconstructions based on $64$ modes/latent features for all approaches. The relative improvement score is calculated in relation to POD performance as baseline.}
    \label{tab:nmss64}
\end{table*}

\subsection{Interpretable latent features}
As described in Sec.~\ref{sec:background}, previous approaches for \gls{cae} based \glspl{rom} proposed interpretable features for high-dimensional dynamical systems. We performed a third set of experiments in which we applied the \gls{smscae} together with the \gls{mdcnnae} and the \gls{hae}. Particularly the \gls{hae} has increased memory and computation time requirements. Hence, we set $\hat{d}=64$ to keep the resource requirements feasible. Because of our observations reported in the last subsection, namely that \gls{smscae} is best trained for $d > \hat{d}$, we use $256$ latent features during training. Nevertheless, the tests are performed on $\hat{d}=64$ features. We also report the number of trainable model parameters along with test reconstruction \gls{nmss} to give an intuition for the model size, cf. Tab.~\ref{tab:nmss64}. For the \gls{pod} baseline, we consider the size of the modes and the average snapshot as training parameters. These conclude to a total of about $266$K values. A considerable small amount compared to the \gls{cae}-based methods. \gls{cae} and \gls{smscae} are within a close range near $20$M parameters. For \gls{mdcnnae} and \gls{hae} the numbers increase by a factor of more than $30$ respectively $63$. The \gls{nmss} shows that all \glspl{cae} except the \gls{hae} clearly outperform the conventional \gls{pod} approach, with relative improvements ranging from $4.9\%$ to $6.4\%$. However, the \gls{hae} leads to a performance decrease ranging from $-15.3$ to $-8.2\%$. \gls{cae}, \gls{mdcnnae} and \gls{smscae} all perform quiet similarly regarding the \gls{nmss}. Thus, there is no visible performance decrease due to the focus on interpretability. Quite the opposite, most results of interpretable methods exceed the conventional \gls{cae} by a small margin.

\begin{figure}[htb]
    \centering
    \includegraphics[width=0.9\linewidth]{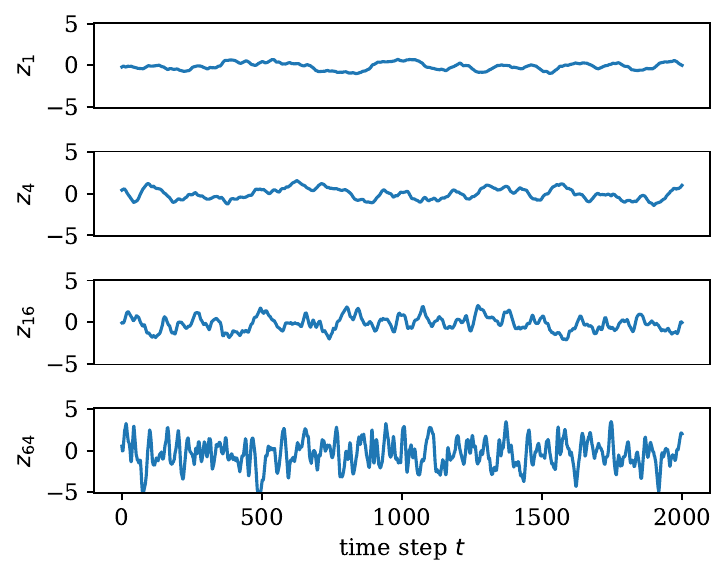}
    \caption{Selection of time-coefficients of POD on the test dataset at $\text{Ra}=9.4\times 10^5$.}
    \label{fig:pod_test_features}
\end{figure}

\begin{figure}[htb]
    \centering
    \includegraphics[width=0.9\linewidth]{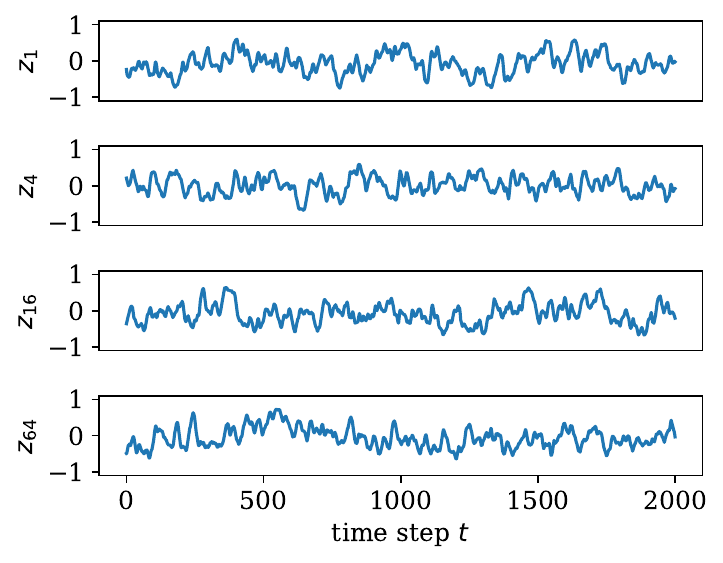}
    \caption{Selection of latent features produced by CAE on the test dataset at $\text{Ra}=9.4\times 10^5$.}
    \label{fig:cae_test_features}
\end{figure}

\begin{figure}[htb]
    \centering
    \includegraphics[width=0.9\linewidth]{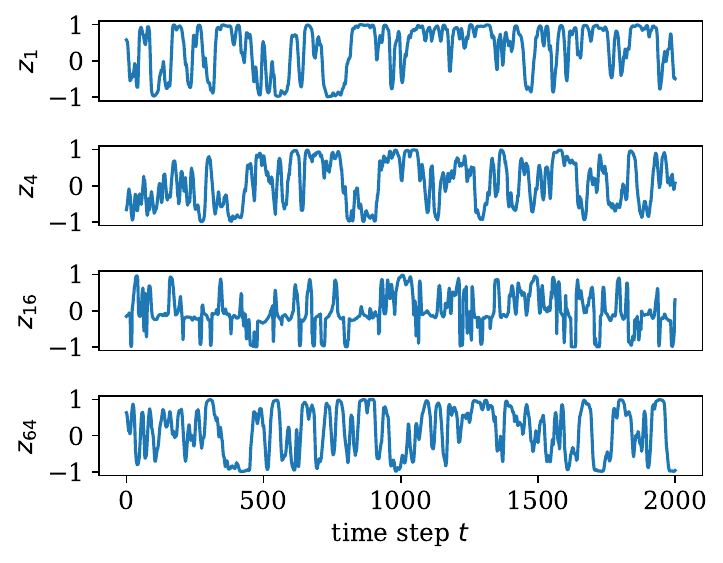}
    \caption{Selection of latent features produced by MD-CNN-AE on the test dataset at $\text{Ra}=9.4\times 10^5$.}
    \label{fig:mdcnnae_test_features}
\end{figure}

\begin{figure}[htb]
    \centering
    \includegraphics[width=0.9\linewidth]{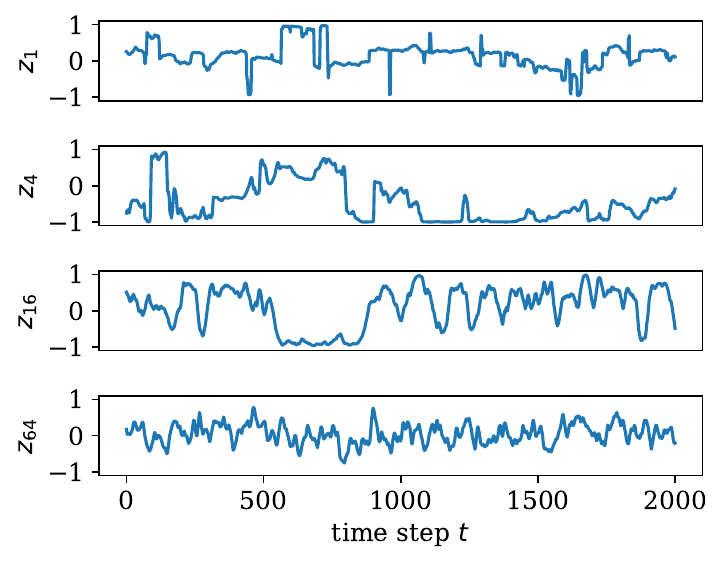}
    \caption{Selection of latent features produced by H-AE on the test dataset at $\text{Ra}=9.4\times 10^5$.}
    \label{fig:hae_test_features}
\end{figure}

\begin{figure}[htb]
    \centering
    \includegraphics[width=0.9\linewidth]{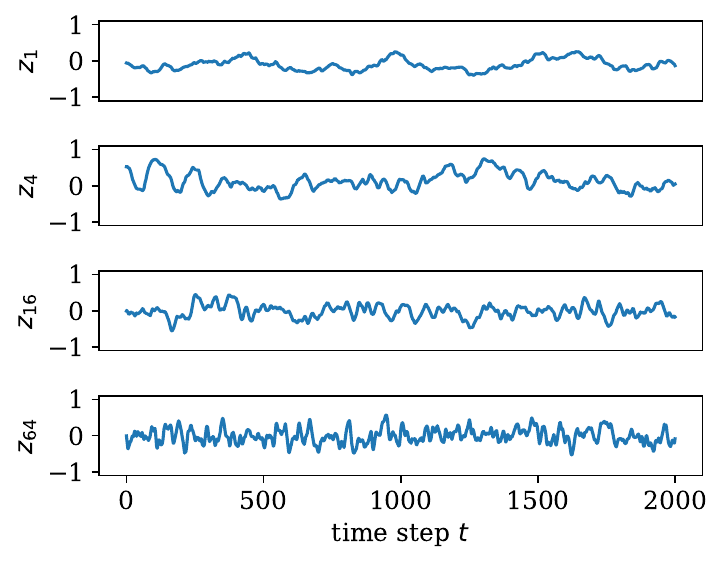}
    \caption{Selection of latent features produced by SMS-CAE on the test dataset at $\text{Ra}=9.4\times 10^5$; compare  to those in previous Figs. \ref{fig:pod_test_features} to \ref{fig:hae_test_features}.}
    \label{fig:smscae_test_features}
\end{figure}

\begin{figure*}[htb]
    \centering
    \includegraphics[width=0.95\linewidth]{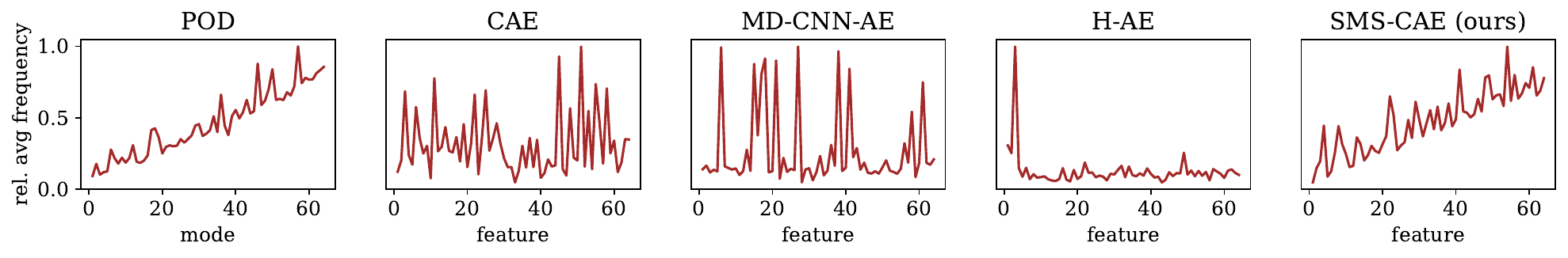}
    \caption{The relative average frequency over the $64$ modes/latent features for POD and different CAE-based ROM methods. The curves for CAE, MD-CNN-AE, H-HAE show no clear trend, but remain relatively constant. POD and SMS-CAE are quiet similar in that the average frequency starts small and increases on average linearly with increasing feature index.}
    \label{fig:rel_avg_freq_per_method}
\end{figure*}

\begin{figure*}[htb]
    \centering
    \includegraphics[width=0.57\linewidth]{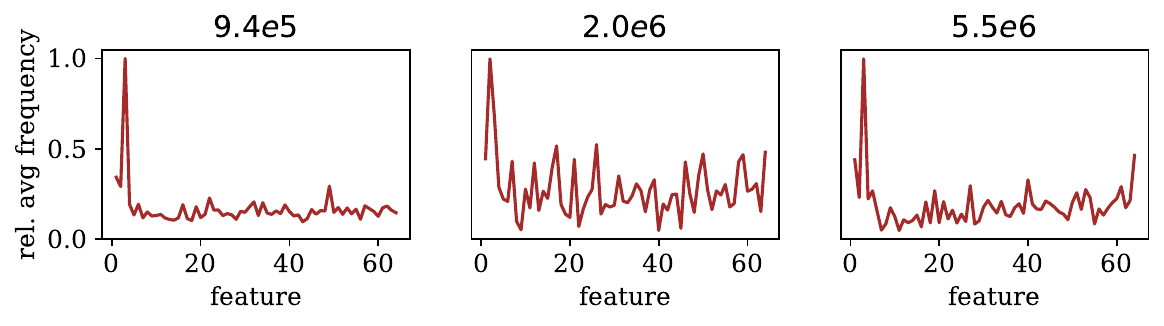}
    \caption{The relative average frequency for H-AE for different Rayleigh numbers $\text{Ra}$ which are given in the title of the panels. We always observe a peak in an early feature.}
    \label{fig:hae_avg_freqs}
\end{figure*}

However, the \gls{nmss} only gives insight on the reconstruction performance, but not the interpretability. Thus, we want to take a closer look on the latent spaces and reconstructions with different $\hat{d} \leq d$. The latent spaces of the conventional \gls{cae} and \gls{mdcnnae} exhibit similar dynamics throughout the different features, cf. Figs.~\ref{fig:cae_test_features} and \ref{fig:mdcnnae_test_features}. \gls{hae} and \gls{smscae} reveal different dynamics for different latent features just like the time-coefficients of \gls{pod} as shown in Figs.~\ref{fig:pod_test_features}, \ref{fig:hae_test_features}, and \ref{fig:smscae_test_features}. It appears that earlier latent features ($z_i$ with smaller $i$) have slower dynamics while at later features the time-series starts to oscillate. The earlier features produced by \gls{hae} remain stable for some time steps before they change abruptly. \gls{smscae} earlier features express a smoother temporal behavior that is more similar to the \gls{pod} coefficients. To further analyse the features dynamics, we calculate the average frequency of feature over time and normalize with the maximum. The result is a curve that displays the course of the relative average frequency over the features which is shown in Fig.~\ref{fig:rel_avg_freq_per_method}. We observe that the frequencies for \gls{cae}, \gls{mdcnnae}, \gls{hae} are rather constant over the features. Only \gls{hae} has a peak at the third feature. Also the test features of the other datasets have a similar peak at an early feature position (cf. Fig.~\ref{fig:hae_avg_freqs}). The frequencies of the \gls{smscae} show a comparable trend to the \gls{pod}. Features with a higher index also show higher average frequencies.

\begin{figure*}[htb]
    \centering
    \includegraphics[width=1.0\linewidth]{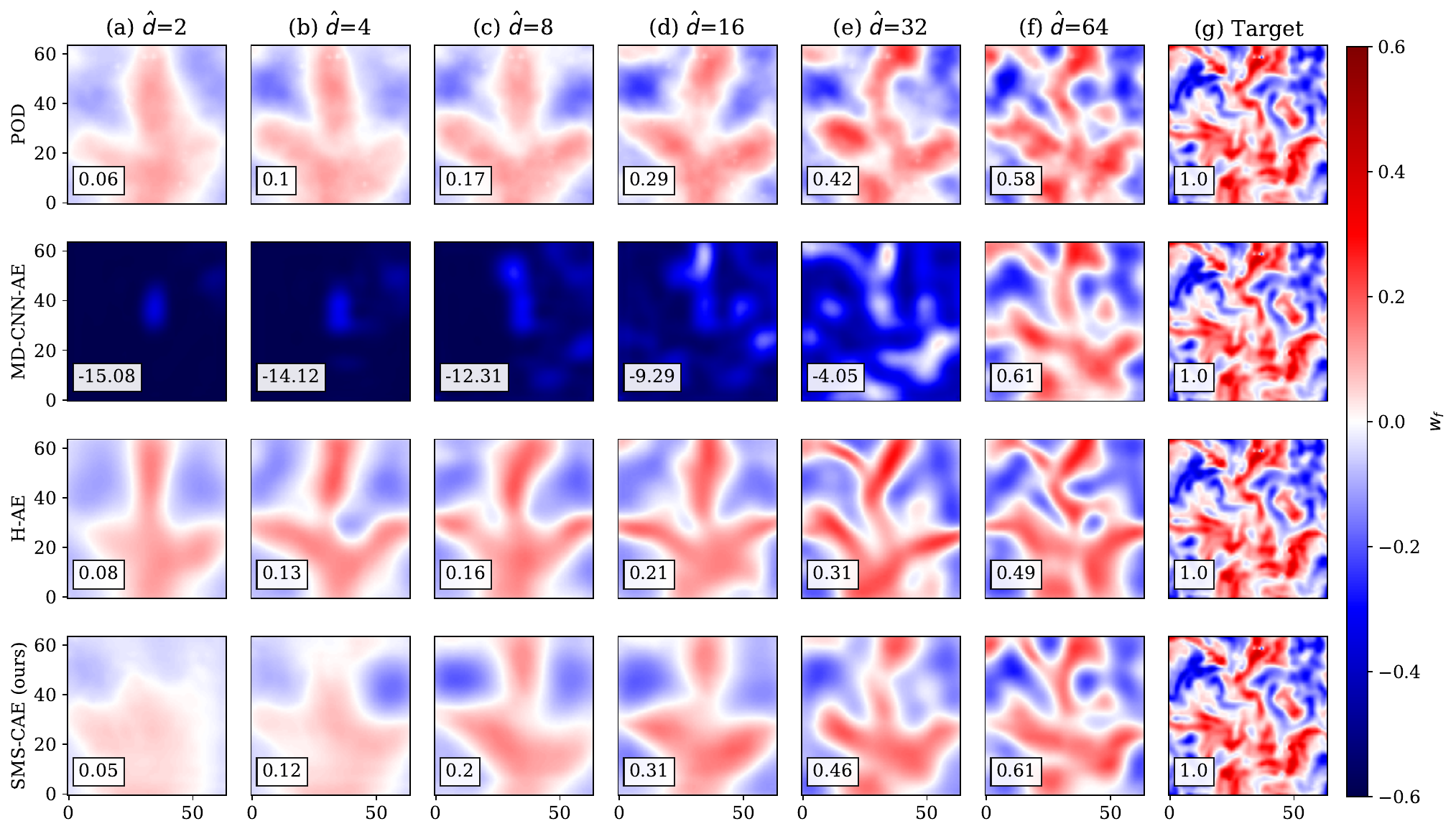}
    \caption{Example test snapshot reconstructions by all tested interpretable methods on different $\hat{d}$. SMS-CAE was trained with $d = 256$ and MD-CNN-AE and H-AE were trained for a $d = 64$. The corresponding average test NMSS is displayed in the lower left corner of each snapshot.}
    \label{fig:all_test_recs}
\end{figure*}

The test snapshot reconstructions by \gls{smscae} and \gls{hae} visually behave similarly to the \gls{pod} reconstructions (cf. Fig.~\ref{fig:all_test_recs}). The autoencoder reconstructions appear smoother and less noisy than those of the \gls{pod}. \gls{smscae} outperforms \gls{pod} for most $\hat{d}$ except the lowest based on test \gls{nmss}. In the region of lower $\hat{d} \leq 4$, \gls{hae} outperforms the \gls{smscae}. However, as $\hat{d}$ increases, \gls{smscae} overtakes while the \gls{nmss} of \gls{hae} falls below that of the \gls{pod} for $\hat{d} \geq 8$. This is due to the hierarchical training scheme of \gls{hae} that primarily favors the lowest $\hat{d}$ in the first stages. In later stages, larger $\hat{d}$ have to cope with features that are optimized for a lower-dimensional task. \gls{mdcnnae} achieves \gls{nmss} similar to \gls{smscae}, but rapidly loses reconstruction quality for $\hat{d} < d$. The \gls{nmss} score drops below $0$ for $\hat{d} \leq 32$, which shows that the kind of interpretability that the method provides is not directly comparable to \gls{pod}. This is due to the applied training procedure and model architecture. They lead to every single features being similarly important for reconstruction. Thus, we cannot select a smallest subset of features that contain most information.

\begin{figure*}[htb]
    \centering
    \includegraphics[width=1.0\linewidth]{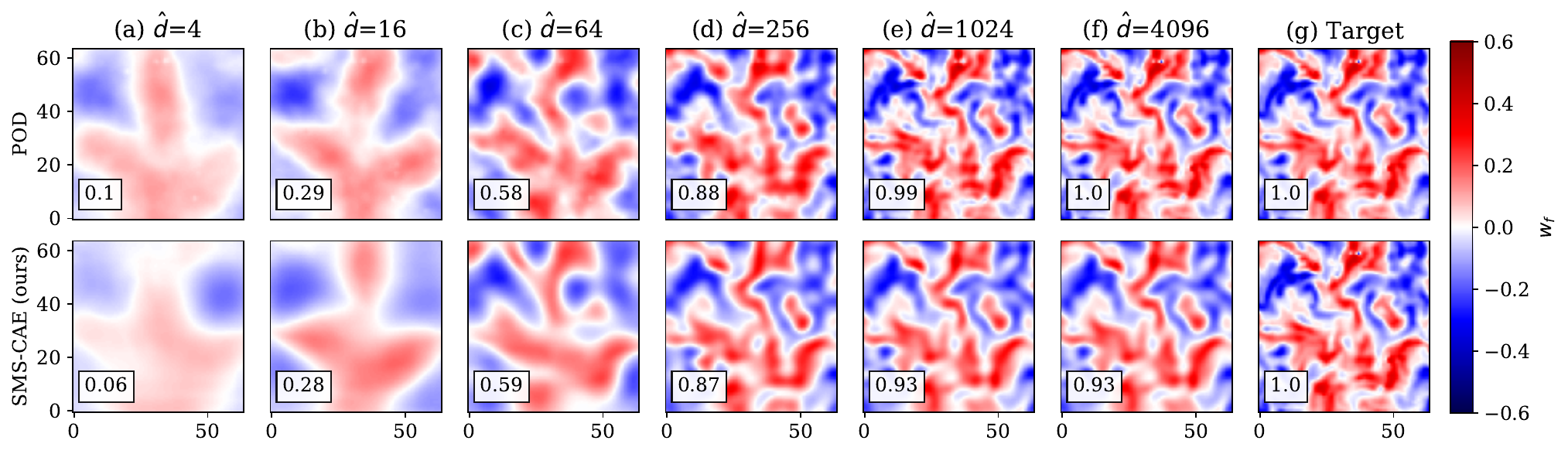}
    \caption{Example test snapshot reconstructions by POD and SMS-CAE based on different $\hat{d}$ while in training $d = 4096$. The corresponding average test NMSS is displayed in the lower left corner of each snapshot.}
    \label{fig:more_test_recs}
\end{figure*}

\subsection{Very low dimensional ROMs}
Our results have shown that \gls{smscae} provides interpretability comparable to \gls{pod} if trained with the \gls{mse} objective function. The reconstruction quality achieved by our method gains a relative improvement of up to $6.4\%$ for $64$ modes. However, as indicated by Figure~\ref{fig:smscae_nmsss}, its reconstruction quality is capped at an \gls{nmss} of about $0.93$. Furthermore, the relative improvement dwindles and falls below that of \gls{pod} as $\hat{d}$ exceeds $\approx10\%$ of the input snapshot size. Therefore, we provide a set of experiments to compare the reconstruction performance of \gls{smscae} and \gls{pod} for very small latent sizes/numbers of modes. Hence, we train the \gls{smscae} with fewer latent features on all three dataset. We report the reconstruction quality for $\hat{d}\in\{1, 2, 4, 8, 16, 32\}$, where $d$ is chosen accordingly during training. We compare the results with the reconstruction quality of \gls{pod} for the same number of modes on the test subsets per dataset (cf. Tab~\ref{tab:smscae_vs_pod}). We report the \gls{nmss} and the relative improvement of \gls{smscae}. The results show a relative improvement ranging from $8.3\%$ and $229.8\%$. Where the most improvement can be observed with smaller \glspl{rom}.

\begin{table*}[htb]
    \centering
    \begin{tabular}{c r | ccr | ccr | ccr }
    \toprule
                                               &           & \multicolumn{3}{c|}{$\text{Ra}=9.4e5$}      & \multicolumn{3}{c|}{$\text{Ra}=2.0e6$} & \multicolumn{3}{c}{$\text{Ra}=5.5e6$}   \\
                                               &           & POD        & SMS-CAE & Rel. impr.    & POD     & SMS-CAE & Rel. impr.  & POD.     & SMS-CAE  & Rel. impr.  \\
    \midrule
    \multirow{6}{*}{\rotatebox{90}{$\hat{d}$}} & $1$       & $0.025$    & $0.052$ & $+104.9\%$ & $0.064$ & $0.104$ & $+61.2\%$   & $0.018$  & $0.058$  & $+229.8\%$   \\
                                               & $2$       & $0.058$    & $0.071$ &  $+22.5\%$ & $0.095$ & $0.126$ & $+33.2\%$   & $0.067$  & $0.093$  &  $+37.4\%$     \\
                                               & $4$       & $0.102$    & $0.145$ &  $+42.9\%$ & $0.145$ & $0.179$ & $+23.6\%$   & $0.113$  & $0.160$  &  $+41.8\%$     \\
                                               & $8$       & $0.172$    & $0.206$ &  $+19.7\%$ & $0.216$ & $0.239$ & $+10.8\%$   & $0.217$  & $0.244$  &  $+12.7\%$      \\
                                               & $16$      & $0.286$    & $0.310$ &   $+8.3\%$ & $0.308$ & $0.335$ & $+8.9\%$    & $0.323$  & $0.368$  &  $+13.7\%$      \\
                                               & $32$      & $0.421$    & $0.459$ &   $+9.1\%$ & $0.432$ & $0.469$ & $+8.6\%$    & $0.440$  & $0.488$  &  $+10.9\%$      \\
    \bottomrule    \end{tabular}
    \caption{NMSS scores and relative improvement on the test data reconstructions with $\hat{d} \in \{1, 2, 4, 8, 16, 32\}$ modes/latent features for all datasets. The relative improvement score is calculated in relation to POD performance as baseline. The results show a clear trend of increased relative improvement for decreased $\hat{d}$.}
    \label{tab:smscae_vs_pod}
\end{table*}

\section{\label{sec:discussion}Discussion}
We developed \gls{smscae}, a novel lightweight multi-scale \gls{cae} that provides interpretability to an extend comparable to \gls{pod}. \gls{smscae} efficiently learns nonlinear \glspl{rom} for high-dimensional and turbulent flows. Our method produces superior interpretability and reconstruction quality compared to existing deep learning-based \gls{rom} approaches. It also requires orders of magnitude less computational resources. We evaluated the performance of our method in a variety of experiments and compared it to related methods.

In our experimental evaluation, we applied a \gls{pod}, a conventional \gls{cae} and three interpretable \glspl{cae} on three turbulent measured \gls{rbc} datasets. We observed that the conventional \gls{cae} produces more precise reconstructions than the \gls{pod} according to \gls{nmss}. The superiority of \gls{cae} is particularly evident when the number of modes is below $10\%$. However, our results indicate that the reconstruction quality of the \gls{cae} solutions saturates at about $93\%$ cumulative energy for our data. Thus, increasing the number of latent features above $400$ for our example case does not lead to recognizable improvements in reconstruction quality with our architecture. We argue that this is due to the data reduction with max-pooling layers. These, are not designed to simply forward information and inevitably lead to loss of information. The loss can no longer be compensated by the succeeding convolutional layers when the corresponding patterns in the data are too noisy. However, since our focus is on data reduction, we consider the advantage in smaller latent spaces to be more relevant to the domain than the disadvantages in ideal reproduction.

The three advanced \glspl{cae} do not only reduce the data, but are designed to produce interpretable features. For \gls{mdcnnae}, each latent feature encodes a separate piece of additive information of the reconstructed snapshot. However, we have no information about the importance of each feature. For \gls{smscae} and \gls{hae}, our results show that their latent representations provide an interpretability similar to \gls{pod}. Thus, we know which latent features encode information about large scale high-energy flow patterns and which about low-energy small scale flow patterns. As a result, we can choose a smallest subset of latent features for reconstruction that preserves as much cumulative energy as possible. The reconstruction quality in terms of test \gls{nmss} shows that \gls{mdcnnae} and \gls{smscae} keep and in some cases even slightly improve the reconstruction quality of the conventional \gls{cae}. However, the \gls{hae} produces clearly poorer reconstructions. We argue that this is due to the curriculum training procedure of the \gls{hae}. Curriculum learning strategies alter the task for the model over the course of training epochs which can stabilize training and speed up convergence \cite{bengio2009curriculum}. The training strategy of the \gls{hae} is designed to find the best sub-model in the current training stage in a greedy way. Thus, the model cannot adapt parameters from earlier stages to be more fit for later ones. The model potentially stays in a local minimum that may turn out to be detrimental later in training. As a result the expressiveness of the whole model is limited. This assumption is supported by Figure~\ref{fig:all_test_recs}, where \gls{hae} provides the strongest reconstructions for small $\hat{d}$, but the weakest for $\hat{d}=d$. Similar behavior was observed in other curriculum learning approaches regarding \gls{rnn} training \cite{teutsch2022flipped}.

With \gls{smscae} we apply a training procedure that considers the reconstruction from different numbers of latent features equally. This helps the model to find a minimum that takes all the different reconstruction scenarios into account. Our results show that \gls{smscae} could not provide the same reconstruction quality as \gls{pod} and \gls{hae} for very small $\hat{d}$ that are also much smaller than $d$. We argue that this is due to the data augmentation, which forces the model to encode the largest structures first \cite{teutsch2024large}. On the one hand, this is also true for the \gls{hae}, but on the other hand, it compensates this due to its training scheme that focuses primarily on the reconstruction with very small $\hat{d}$.

To produce better results for very small $\hat{d}$ with \gls{smscae}, the $d$ should be choosen accordingly during training. We conducted another set of comparative experiments for \gls{pod} and \gls{smscae} for reconstructions on very small numbers of modes. We observed that the relative improvement of our method, increases greatly as the number of modes/features decreases. We argue, that the nonlinear advantage of \gls{smscae} develops mostly in the first most informative latent features.

In the feature analysis we observed that \gls{pod} and \gls{smscae} features behaved similarly. In both cases, the relative average frequency increased with the feature index. Thus, a feature with a smaller index holds information about slower temporal dynamics whereas features with a larger index take care of the faster dynamics. \gls{cae} and \gls{mdcnnae} do not show a similar trend and cannot be interpreted this way. The \gls{hae} features for all tested datasets have shown a clear frequency peak at the second or third feature. Apart from that, we cannot deduce a trend in the feature frequencies.

Besides reconstruction quality and interpretability, our results also reveal a large margin between the different approaches' model sizes regarding trainable parameters. While \gls{smscae} only weighs in at a minimum increase by about $10\%$, \gls{mdcnnae} and \gls{hae} increase the number of parameters by $300\%$ and $630\%$ compared to the base \gls{cae}. For the \gls{smscae}, this is due to an increased latent size during training. Our results show that a $2$ to $4$ times larger latent size $d$ leads to improved reconstruction performance with the target latent size $\hat{d}$. Thus, during inference we can discard the additional parameters and the model size is equal to the conventional \gls{cae}. \gls{mdcnnae} uses multiple sub-decoders which leads to the immense increase in trainable parameters. For the \gls{hae} this increase is even higher, because it uses multiple sub-encoders and multiple sub-decoders. This increased number of parameters alone leads not only to larger memory requirements, but also slows down training speed and inference in practice. Additionally, for the \gls{hae}, hierarchical training procedure requires to let each pair of sub-encoder and sub-decoder converge before the next pair is trained. This adds a number of epochs at each hierarchical stage to ensure the convergence. During these epochs, the model does not improve and their amount scales linearly with the number of sub-encoder-decoder pairs.

\section{\label{sec:conclusion}Conclusion}
In this paper, we studied interpretable \gls{cae}-based \gls{rom} approaches for high-dimensional turbulent flows. We contextualized them with the widely used \gls{pod} and compared their output in terms of reconstruction quality and interpretability. To assess the reconstruction quality of \gls{pod} and \gls{cae}, we used the \gls{nmss} as performance metric. It behaved similarly to the cumulative energy curve often used in the context of \gls{pod} and can be applied to \gls{cae} reconstructions as well. As \gls{cae}-based \glspl{rom}, we used the \gls{mdcnnae} and the \gls{hae}, both of which use an adapted architecture to improve the interpretability of the latent representations. As a new approach, we proposed the \gls{smscae}. It adds a constrained dropout layer to any existing conventional \gls{cae}. It controls the importance of each feature for reconstruction. In our experiments, we applied the different methods for three measured 2-dimensional \gls{rbc} datasets. All \gls{cae}-based methods except for \gls{hae} beat \gls{pod} in terms of compression and reconstruction quality. Our results show that \gls{smscae} offers interpretability similar to \gls{pod}. It delivers excellent reconstruction performance, with relative improvements over \gls{pod} ranging from $4.9\%$ to $229.8\%$ depending on dataset and number of modes. \gls{smscae} is lightweight and easier to train compared to \gls{mdcnnae} and \gls{hae}. \gls{smscae} outperforms existing methods in terms of reconstruction quality and interpretability.

\gls{smscae} not only achieves convincing results, but also motivates future research that opens the black box of deep learning based methods a little further. With \gls{smscae} as a starting point this means developing and testing constrained dropout with different objective functions to shift the feature focus, and investigating how other types of models that are not autoencoders behave when forced to prioritize their latent features. For deep learning aided fluid mechanics, this means combining the technique with physics-informed approaches, with the ultimate goal of truly interpreting features in terms of absolute real-world scales.

\begin{acknowledgments}
The work is supported by the Carl-Zeiss-Foundation within project no. P2018-02-001 “Deep Turb - Deep Learning in and of Turbulence" and project no. P2022-08-006 "PollenNet", and the Federal Ministry for the Environment, Nature Conservation, Nuclear Safety and Consumer Protection (BMUV) project no. 67KI2086A "Natura Incognita".
\end{acknowledgments}

\section*{Data availability}
The data that supports the findings of this paper are openly available on \href{https://doi.org/10.7910/DVN/2M8RKD}{Dataverse}. The reproduction code is publicly available via \href{https://doi.org/10.5281/zenodo.13960758}{Zenodo}.



\section*{Bibliography}
\bibliography{aipsamp}

\begin{thebibliography}{46}%
\makeatletter
\providecommand \@ifxundefined [1]{%
 \@ifx{#1\undefined}
}%
\providecommand \@ifnum [1]{%
 \ifnum #1\expandafter \@firstoftwo
 \else \expandafter \@secondoftwo
 \fi
}%
\providecommand \@ifx [1]{%
 \ifx #1\expandafter \@firstoftwo
 \else \expandafter \@secondoftwo
 \fi
}%
\providecommand \natexlab [1]{#1}%
\providecommand \enquote  [1]{``#1''}%
\providecommand \bibnamefont  [1]{#1}%
\providecommand \bibfnamefont [1]{#1}%
\providecommand \citenamefont [1]{#1}%
\providecommand \href@noop [0]{\@secondoftwo}%
\providecommand \href [0]{\begingroup \@sanitize@url \@href}%
\providecommand \@href[1]{\@@startlink{#1}\@@href}%
\providecommand \@@href[1]{\endgroup#1\@@endlink}%
\providecommand \@sanitize@url [0]{\catcode `\\12\catcode `\$12\catcode
  `\&12\catcode `\#12\catcode `\^12\catcode `\_12\catcode `\%12\relax}%
\providecommand \@@startlink[1]{}%
\providecommand \@@endlink[0]{}%
\providecommand \url  [0]{\begingroup\@sanitize@url \@url }%
\providecommand \@url [1]{\endgroup\@href {#1}{\urlprefix }}%
\providecommand \urlprefix  [0]{URL }%
\providecommand \Eprint [0]{\href }%
\providecommand \doibase [0]{http://dx.doi.org/}%
\providecommand \selectlanguage [0]{\@gobble}%
\providecommand \bibinfo  [0]{\@secondoftwo}%
\providecommand \bibfield  [0]{\@secondoftwo}%
\providecommand \translation [1]{[#1]}%
\providecommand \BibitemOpen [0]{}%
\providecommand \bibitemStop [0]{}%
\providecommand \bibitemNoStop [0]{.\EOS\space}%
\providecommand \EOS [0]{\spacefactor3000\relax}%
\providecommand \BibitemShut  [1]{\csname bibitem#1\endcsname}%
\let\auto@bib@innerbib\@empty
\bibitem [{\citenamefont {Brunton}, \citenamefont {Noack},\ and\ \citenamefont
  {Koumoutsakos}(2020)}]{brunton2020machine}%
  \BibitemOpen
  \bibfield  {author} {\bibinfo {author} {\bibfnamefont {S.~L.}\ \bibnamefont
  {Brunton}}, \bibinfo {author} {\bibfnamefont {B.~R.}\ \bibnamefont {Noack}},
  \ and\ \bibinfo {author} {\bibfnamefont {P.}~\bibnamefont {Koumoutsakos}},\
  }\bibfield  {title} {\enquote {\bibinfo {title} {Machine {L}earning for
  {F}luid {M}echanics},}\ }\href@noop {} {\bibfield  {journal} {\bibinfo
  {journal} {Annual Review of Fluid Mechanics}\ }\textbf {\bibinfo {volume}
  {52}},\ \bibinfo {pages} {477--508} (\bibinfo {year} {2020})}\BibitemShut
  {NoStop}%
\bibitem [{\citenamefont {Lu}\ \emph {et~al.}(2017)\citenamefont {Lu},
  \citenamefont {Pathak}, \citenamefont {Hunt}, \citenamefont {Girvan},
  \citenamefont {Brockett},\ and\ \citenamefont {Ott}}]{lu2017reservoir}%
  \BibitemOpen
  \bibfield  {author} {\bibinfo {author} {\bibfnamefont {Z.}~\bibnamefont
  {Lu}}, \bibinfo {author} {\bibfnamefont {J.}~\bibnamefont {Pathak}}, \bibinfo
  {author} {\bibfnamefont {B.}~\bibnamefont {Hunt}}, \bibinfo {author}
  {\bibfnamefont {M.}~\bibnamefont {Girvan}}, \bibinfo {author} {\bibfnamefont
  {R.}~\bibnamefont {Brockett}}, \ and\ \bibinfo {author} {\bibfnamefont
  {E.}~\bibnamefont {Ott}},\ }\bibfield  {title} {\enquote {\bibinfo {title}
  {{Reservoir observers: {M}odel-free inference of unmeasured variables in
  chaotic systems}},}\ }\href {\doibase 10.1063/1.4979665} {\bibfield
  {journal} {\bibinfo  {journal} {Chaos: An Interdisciplinary Journal of
  Nonlinear Science}\ }\textbf {\bibinfo {volume} {27}},\ \bibinfo {pages}
  {041102} (\bibinfo {year} {2017})},\ \Eprint
  {http://arxiv.org/abs/https://pubs.aip.org/aip/cha/article-pdf/doi/10.1063/1.4979665/10313493/041102\_1\_online.pdf}
  {https://pubs.aip.org/aip/cha/article-pdf/doi/10.1063/1.4979665/10313493/041102\_1\_online.pdf}
  \BibitemShut {NoStop}%
\bibitem [{\citenamefont {Raissi}, \citenamefont {Yazdani},\ and\ \citenamefont
  {Karniadakis}(2020)}]{raissi2020hidden}%
  \BibitemOpen
  \bibfield  {author} {\bibinfo {author} {\bibfnamefont {M.}~\bibnamefont
  {Raissi}}, \bibinfo {author} {\bibfnamefont {A.}~\bibnamefont {Yazdani}}, \
  and\ \bibinfo {author} {\bibfnamefont {G.~E.}\ \bibnamefont {Karniadakis}},\
  }\bibfield  {title} {\enquote {\bibinfo {title} {Hidden fluid mechanics:
  {L}earning velocity and pressure fields from flow visualizations},}\
  }\href@noop {} {\bibfield  {journal} {\bibinfo  {journal} {Science}\ }\textbf
  {\bibinfo {volume} {367}},\ \bibinfo {pages} {1026--1030} (\bibinfo {year}
  {2020})}\BibitemShut {NoStop}%
\bibitem [{\citenamefont {Tian}\ \emph {et~al.}(2023)\citenamefont {Tian},
  \citenamefont {Woodward}, \citenamefont {Stepanov}, \citenamefont {Fryer},
  \citenamefont {Hyett}, \citenamefont {Livescu},\ and\ \citenamefont
  {Chertkov}}]{tian2023lagrangian}%
  \BibitemOpen
  \bibfield  {author} {\bibinfo {author} {\bibfnamefont {Y.}~\bibnamefont
  {Tian}}, \bibinfo {author} {\bibfnamefont {M.}~\bibnamefont {Woodward}},
  \bibinfo {author} {\bibfnamefont {M.}~\bibnamefont {Stepanov}}, \bibinfo
  {author} {\bibfnamefont {C.}~\bibnamefont {Fryer}}, \bibinfo {author}
  {\bibfnamefont {C.}~\bibnamefont {Hyett}}, \bibinfo {author} {\bibfnamefont
  {D.}~\bibnamefont {Livescu}}, \ and\ \bibinfo {author} {\bibfnamefont
  {M.}~\bibnamefont {Chertkov}},\ }\bibfield  {title} {\enquote {\bibinfo
  {title} {Lagrangian large eddy simulations via physics-informed machine
  learning},}\ }\href@noop {} {\bibfield  {journal} {\bibinfo  {journal}
  {Proceedings of the National Academy of Sciences}\ }\textbf {\bibinfo
  {volume} {120}},\ \bibinfo {pages} {e2213638120} (\bibinfo {year}
  {2023})}\BibitemShut {NoStop}%
\bibitem [{\citenamefont {Fonda}\ \emph {et~al.}(2019)\citenamefont {Fonda},
  \citenamefont {Pandey}, \citenamefont {Schumacher},\ and\ \citenamefont
  {Sreenivasan}}]{fonda2019deep}%
  \BibitemOpen
  \bibfield  {author} {\bibinfo {author} {\bibfnamefont {E.}~\bibnamefont
  {Fonda}}, \bibinfo {author} {\bibfnamefont {A.}~\bibnamefont {Pandey}},
  \bibinfo {author} {\bibfnamefont {J.}~\bibnamefont {Schumacher}}, \ and\
  \bibinfo {author} {\bibfnamefont {K.~R.}\ \bibnamefont {Sreenivasan}},\
  }\bibfield  {title} {\enquote {\bibinfo {title} {Deep learning in turbulent
  convection networks},}\ }\href@noop {} {\bibfield  {journal} {\bibinfo
  {journal} {Proceedings of the National Academy of Sciences}\ }\textbf
  {\bibinfo {volume} {116}},\ \bibinfo {pages} {8667--8672} (\bibinfo {year}
  {2019})}\BibitemShut {NoStop}%
\bibitem [{\citenamefont {Pandey}\ and\ \citenamefont
  {Schumacher}(2020)}]{pandey2020reservoir}%
  \BibitemOpen
  \bibfield  {author} {\bibinfo {author} {\bibfnamefont {S.}~\bibnamefont
  {Pandey}}\ and\ \bibinfo {author} {\bibfnamefont {J.}~\bibnamefont
  {Schumacher}},\ }\bibfield  {title} {\enquote {\bibinfo {title} {Reservoir
  computing model of two-dimensional turbulent convection},}\ }\href@noop {}
  {\bibfield  {journal} {\bibinfo  {journal} {Physical Review Fluids}\ }\textbf
  {\bibinfo {volume} {5}},\ \bibinfo {pages} {113506} (\bibinfo {year}
  {2020})}\BibitemShut {NoStop}%
\bibitem [{\citenamefont {Pfeffer}, \citenamefont {Heyder},\ and\ \citenamefont
  {Schumacher}(2022)}]{pfeffer2022hybrid}%
  \BibitemOpen
  \bibfield  {author} {\bibinfo {author} {\bibfnamefont {P.}~\bibnamefont
  {Pfeffer}}, \bibinfo {author} {\bibfnamefont {F.}~\bibnamefont {Heyder}}, \
  and\ \bibinfo {author} {\bibfnamefont {J.}~\bibnamefont {Schumacher}},\
  }\bibfield  {title} {\enquote {\bibinfo {title} {Hybrid quantum-classical
  reservoir computing of thermal convection flow},}\ }\href@noop {} {\bibfield
  {journal} {\bibinfo  {journal} {Physical Review Research}\ }\textbf {\bibinfo
  {volume} {4}},\ \bibinfo {pages} {033176} (\bibinfo {year}
  {2022})}\BibitemShut {NoStop}%
\bibitem [{\citenamefont {Teutsch}\ \emph {et~al.}(2023)\citenamefont
  {Teutsch}, \citenamefont {K{\"a}ufer}, \citenamefont {M{\"a}der},\ and\
  \citenamefont {Cierpka}}]{teutsch2023data}%
  \BibitemOpen
  \bibfield  {author} {\bibinfo {author} {\bibfnamefont {P.}~\bibnamefont
  {Teutsch}}, \bibinfo {author} {\bibfnamefont {T.}~\bibnamefont {K{\"a}ufer}},
  \bibinfo {author} {\bibfnamefont {P.}~\bibnamefont {M{\"a}der}}, \ and\
  \bibinfo {author} {\bibfnamefont {C.}~\bibnamefont {Cierpka}},\ }\bibfield
  {title} {\enquote {\bibinfo {title} {Data-driven estimation of scalar
  quantities from planar velocity measurements by deep learning applied to
  temperature in thermal convection},}\ }\href@noop {} {\bibfield  {journal}
  {\bibinfo  {journal} {Experiments in Fluids}\ }\textbf {\bibinfo {volume}
  {64}},\ \bibinfo {pages} {191} (\bibinfo {year} {2023})}\BibitemShut
  {NoStop}%
\bibitem [{\citenamefont {Biferale}, \citenamefont {Buzzicotti},\ and\
  \citenamefont {Cencini}(2023)}]{biferale2023topical}%
  \BibitemOpen
  \bibfield  {author} {\bibinfo {author} {\bibfnamefont {L.}~\bibnamefont
  {Biferale}}, \bibinfo {author} {\bibfnamefont {M.}~\bibnamefont
  {Buzzicotti}}, \ and\ \bibinfo {author} {\bibfnamefont {M.}~\bibnamefont
  {Cencini}},\ }\bibfield  {title} {\enquote {\bibinfo {title} {Topical issue
  on quantitative {AI} in complex fluids and complex flows: challenges and
  benchmarks},}\ }\href@noop {} {\bibfield  {journal} {\bibinfo  {journal} {The
  European Physical Journal E}\ }\textbf {\bibinfo {volume} {46}},\ \bibinfo
  {pages} {102} (\bibinfo {year} {2023})}\BibitemShut {NoStop}%
\bibitem [{\citenamefont {Li}\ \emph {et~al.}(2024)\citenamefont {Li},
  \citenamefont {Biferale}, \citenamefont {Bonaccorso}, \citenamefont
  {Scarpolini},\ and\ \citenamefont {Buzzicotti}}]{li2024synthetic}%
  \BibitemOpen
  \bibfield  {author} {\bibinfo {author} {\bibfnamefont {T.}~\bibnamefont
  {Li}}, \bibinfo {author} {\bibfnamefont {L.}~\bibnamefont {Biferale}},
  \bibinfo {author} {\bibfnamefont {F.}~\bibnamefont {Bonaccorso}}, \bibinfo
  {author} {\bibfnamefont {M.~A.}\ \bibnamefont {Scarpolini}}, \ and\ \bibinfo
  {author} {\bibfnamefont {M.}~\bibnamefont {Buzzicotti}},\ }\bibfield  {title}
  {\enquote {\bibinfo {title} {Synthetic {L}agrangian turbulence by generative
  diffusion models},}\ }\href@noop {} {\bibfield  {journal} {\bibinfo
  {journal} {Nature Machine Intelligence}\ ,\ \bibinfo {pages} {1--11}}
  (\bibinfo {year} {2024})}\BibitemShut {NoStop}%
\bibitem [{\citenamefont {Salim}, \citenamefont {Burkhart},\ and\ \citenamefont
  {Sondak}(2024)}]{salim2024extending}%
  \BibitemOpen
  \bibfield  {author} {\bibinfo {author} {\bibfnamefont {D.~M.}\ \bibnamefont
  {Salim}}, \bibinfo {author} {\bibfnamefont {B.}~\bibnamefont {Burkhart}}, \
  and\ \bibinfo {author} {\bibfnamefont {D.}~\bibnamefont {Sondak}},\
  }\bibfield  {title} {\enquote {\bibinfo {title} {Extending a
  {P}hysics-informed {M}achine-learning {N}etwork for {S}uperresolution
  {S}tudies of {R}ayleigh--{B}{\'e}nard {C}onvection},}\ }\href@noop {}
  {\bibfield  {journal} {\bibinfo  {journal} {The Astrophysical Journal}\
  }\textbf {\bibinfo {volume} {964}},\ \bibinfo {pages} {2} (\bibinfo {year}
  {2024})}\BibitemShut {NoStop}%
\bibitem [{\citenamefont {Vlachas}\ \emph {et~al.}(2018)\citenamefont
  {Vlachas}, \citenamefont {Byeon}, \citenamefont {Wan}, \citenamefont
  {Sapsis},\ and\ \citenamefont {Koumoutsakos}}]{vlachas2018data}%
  \BibitemOpen
  \bibfield  {author} {\bibinfo {author} {\bibfnamefont {P.~R.}\ \bibnamefont
  {Vlachas}}, \bibinfo {author} {\bibfnamefont {W.}~\bibnamefont {Byeon}},
  \bibinfo {author} {\bibfnamefont {Z.~Y.}\ \bibnamefont {Wan}}, \bibinfo
  {author} {\bibfnamefont {T.~P.}\ \bibnamefont {Sapsis}}, \ and\ \bibinfo
  {author} {\bibfnamefont {P.}~\bibnamefont {Koumoutsakos}},\ }\bibfield
  {title} {\enquote {\bibinfo {title} {Data-driven forecasting of
  high-dimensional chaotic systems with long short-term memory networks},}\
  }\href@noop {} {\bibfield  {journal} {\bibinfo  {journal} {Proceedings of the
  Royal Society A: Mathematical, Physical and Engineering Sciences}\ }\textbf
  {\bibinfo {volume} {474}},\ \bibinfo {pages} {20170844} (\bibinfo {year}
  {2018})}\BibitemShut {NoStop}%
\bibitem [{\citenamefont {Maulik}, \citenamefont {Lusch},\ and\ \citenamefont
  {Balaprakash}(2021)}]{maulik2021reduced}%
  \BibitemOpen
  \bibfield  {author} {\bibinfo {author} {\bibfnamefont {R.}~\bibnamefont
  {Maulik}}, \bibinfo {author} {\bibfnamefont {B.}~\bibnamefont {Lusch}}, \
  and\ \bibinfo {author} {\bibfnamefont {P.}~\bibnamefont {Balaprakash}},\
  }\bibfield  {title} {\enquote {\bibinfo {title} {{Reduced-order modeling of
  advection-dominated systems with recurrent neural networks and convolutional
  autoencoders}},}\ }\href {\doibase 10.1063/5.0039986} {\bibfield  {journal}
  {\bibinfo  {journal} {Physics of Fluids}\ }\textbf {\bibinfo {volume} {33}},\
  \bibinfo {pages} {037106} (\bibinfo {year} {2021})},\ \Eprint
  {http://arxiv.org/abs/https://pubs.aip.org/aip/pof/article-pdf/doi/10.1063/5.0039986/15666155/037106\_1\_online.pdf}
  {https://pubs.aip.org/aip/pof/article-pdf/doi/10.1063/5.0039986/15666155/037106\_1\_online.pdf}
  \BibitemShut {NoStop}%
\bibitem [{\citenamefont {Phillips}\ \emph {et~al.}(2021)\citenamefont
  {Phillips}, \citenamefont {Heaney}, \citenamefont {Smith},\ and\
  \citenamefont {Pain}}]{phillips2021autoencoder}%
  \BibitemOpen
  \bibfield  {author} {\bibinfo {author} {\bibfnamefont {T.~R.}\ \bibnamefont
  {Phillips}}, \bibinfo {author} {\bibfnamefont {C.~E.}\ \bibnamefont
  {Heaney}}, \bibinfo {author} {\bibfnamefont {P.~N.}\ \bibnamefont {Smith}}, \
  and\ \bibinfo {author} {\bibfnamefont {C.~C.}\ \bibnamefont {Pain}},\
  }\bibfield  {title} {\enquote {\bibinfo {title} {An autoencoder-based
  reduced-order model for eigenvalue problems with application to neutron
  diffusion},}\ }\href@noop {} {\bibfield  {journal} {\bibinfo  {journal}
  {International Journal for Numerical Methods in Engineering}\ }\textbf
  {\bibinfo {volume} {122}},\ \bibinfo {pages} {3780--3811} (\bibinfo {year}
  {2021})}\BibitemShut {NoStop}%
\bibitem [{\citenamefont {Pandey}\ \emph {et~al.}(2022)\citenamefont {Pandey},
  \citenamefont {Teutsch}, \citenamefont {M{\"a}der},\ and\ \citenamefont
  {Schumacher}}]{pandey2022direct}%
  \BibitemOpen
  \bibfield  {author} {\bibinfo {author} {\bibfnamefont {S.}~\bibnamefont
  {Pandey}}, \bibinfo {author} {\bibfnamefont {P.}~\bibnamefont {Teutsch}},
  \bibinfo {author} {\bibfnamefont {P.}~\bibnamefont {M{\"a}der}}, \ and\
  \bibinfo {author} {\bibfnamefont {J.}~\bibnamefont {Schumacher}},\ }\bibfield
   {title} {\enquote {\bibinfo {title} {Direct data-driven forecast of local
  turbulent heat flux in {R}ayleigh--{B}{\'e}nard convection},}\ }\href@noop {}
  {\bibfield  {journal} {\bibinfo  {journal} {Physics of Fluids}\ }\textbf
  {\bibinfo {volume} {34}} (\bibinfo {year} {2022})}\BibitemShut {NoStop}%
\bibitem [{\citenamefont {Yongho}\ and\ \citenamefont
  {Heiland}(2023)}]{yongho2023convolutional}%
  \BibitemOpen
  \bibfield  {author} {\bibinfo {author} {\bibfnamefont {K.}~\bibnamefont
  {Yongho}}\ and\ \bibinfo {author} {\bibfnamefont {J.}~\bibnamefont
  {Heiland}},\ }\bibfield  {title} {\enquote {\bibinfo {title} {Convolutional
  autoencoders, clustering and {POD} for low-dimensional parametrization of
  {N}avier-{S}tokes equations},}\ }\href@noop {} {\bibfield  {journal}
  {\bibinfo  {journal} {arXiv preprint arXiv:2302.01278}\ } (\bibinfo {year}
  {2023})}\BibitemShut {NoStop}%
\bibitem [{\citenamefont {Berkooz}, \citenamefont {Holmes},\ and\ \citenamefont
  {Lumley}(1993)}]{berkooz1993proper}%
  \BibitemOpen
  \bibfield  {author} {\bibinfo {author} {\bibfnamefont {G.}~\bibnamefont
  {Berkooz}}, \bibinfo {author} {\bibfnamefont {P.}~\bibnamefont {Holmes}}, \
  and\ \bibinfo {author} {\bibfnamefont {J.~L.}\ \bibnamefont {Lumley}},\
  }\bibfield  {title} {\enquote {\bibinfo {title} {The {P}roper {O}rthogonal
  {D}ecomposition in the analysis of turbulent flows},}\ }\href@noop {}
  {\bibfield  {journal} {\bibinfo  {journal} {Annual Review of Fluid
  Mechanics}\ }\textbf {\bibinfo {volume} {25}},\ \bibinfo {pages} {539--575}
  (\bibinfo {year} {1993})}\BibitemShut {NoStop}%
\bibitem [{\citenamefont {Weiss}(2019)}]{weiss2019tutorial}%
  \BibitemOpen
  \bibfield  {author} {\bibinfo {author} {\bibfnamefont {J.}~\bibnamefont
  {Weiss}},\ }\bibfield  {title} {\enquote {\bibinfo {title} {A tutorial on the
  {P}roper {O}rthogonal {D}ecomposition},}\ }\bibfield  {booktitle} {\emph
  {\bibinfo {booktitle} {AIAA Aviation 2019 Forum}},\ }\href {\doibase
  10.2514/6.2019-3333} {\bibfield  {journal} {\bibinfo  {journal} {AIAA
  Aviation Forum}\ ,\ \bibinfo {pages} {3333}} (\bibinfo {year} {2019})},\
  \Eprint
  {http://arxiv.org/abs/https://arc.aiaa.org/doi/pdf/10.2514/6.2019-3333}
  {https://arc.aiaa.org/doi/pdf/10.2514/6.2019-3333} \BibitemShut {NoStop}%
\bibitem [{\citenamefont {Chatterjee}(2000)}]{chatterjee2000introduction}%
  \BibitemOpen
  \bibfield  {author} {\bibinfo {author} {\bibfnamefont {A.}~\bibnamefont
  {Chatterjee}},\ }\bibfield  {title} {\enquote {\bibinfo {title} {An
  introduction to the proper orthogonal decomposition},}\ }\href@noop {}
  {\bibfield  {journal} {\bibinfo  {journal} {Current Science}\ ,\ \bibinfo
  {pages} {808--817}} (\bibinfo {year} {2000})}\BibitemShut {NoStop}%
\bibitem [{\citenamefont {Gonzalez}\ and\ \citenamefont
  {Balajewicz}(2018)}]{gonzalez2018deep}%
  \BibitemOpen
  \bibfield  {author} {\bibinfo {author} {\bibfnamefont {F.~J.}\ \bibnamefont
  {Gonzalez}}\ and\ \bibinfo {author} {\bibfnamefont {M.}~\bibnamefont
  {Balajewicz}},\ }\bibfield  {title} {\enquote {\bibinfo {title} {Deep
  convolutional recurrent autoencoders for learning low-dimensional feature
  dynamics of fluid systems},}\ }\href@noop {} {\bibfield  {journal} {\bibinfo
  {journal} {arXiv preprint arXiv:1808.01346}\ } (\bibinfo {year}
  {2018})}\BibitemShut {NoStop}%
\bibitem [{\citenamefont {Ahmed}\ \emph {et~al.}(2021)\citenamefont {Ahmed},
  \citenamefont {San}, \citenamefont {Rasheed},\ and\ \citenamefont
  {Iliescu}}]{ahmed2021nonlinear}%
  \BibitemOpen
  \bibfield  {author} {\bibinfo {author} {\bibfnamefont {S.~E.}\ \bibnamefont
  {Ahmed}}, \bibinfo {author} {\bibfnamefont {O.}~\bibnamefont {San}}, \bibinfo
  {author} {\bibfnamefont {A.}~\bibnamefont {Rasheed}}, \ and\ \bibinfo
  {author} {\bibfnamefont {T.}~\bibnamefont {Iliescu}},\ }\bibfield  {title}
  {\enquote {\bibinfo {title} {{Nonlinear proper orthogonal decomposition for
  convection-dominated flows}},}\ }\href {\doibase 10.1063/5.0074310}
  {\bibfield  {journal} {\bibinfo  {journal} {Physics of Fluids}\ }\textbf
  {\bibinfo {volume} {33}},\ \bibinfo {pages} {121702} (\bibinfo {year}
  {2021})},\ \Eprint
  {http://arxiv.org/abs/https://pubs.aip.org/aip/pof/article-pdf/doi/10.1063/5.0074310/15857652/121702\_1\_online.pdf}
  {https://pubs.aip.org/aip/pof/article-pdf/doi/10.1063/5.0074310/15857652/121702\_1\_online.pdf}
  \BibitemShut {NoStop}%
\bibitem [{\citenamefont {Obayashi}\ \emph {et~al.}(2021)\citenamefont
  {Obayashi}, \citenamefont {Aono}, \citenamefont {Tatsukawa},\ and\
  \citenamefont {Fujii}}]{obayashi2021feature}%
  \BibitemOpen
  \bibfield  {author} {\bibinfo {author} {\bibfnamefont {W.}~\bibnamefont
  {Obayashi}}, \bibinfo {author} {\bibfnamefont {H.}~\bibnamefont {Aono}},
  \bibinfo {author} {\bibfnamefont {T.}~\bibnamefont {Tatsukawa}}, \ and\
  \bibinfo {author} {\bibfnamefont {K.}~\bibnamefont {Fujii}},\ }\bibfield
  {title} {\enquote {\bibinfo {title} {Feature extraction of fields of fluid
  dynamics data using sparse convolutional autoencoder},}\ }\href@noop {}
  {\bibfield  {journal} {\bibinfo  {journal} {AIP Advances}\ }\textbf {\bibinfo
  {volume} {11}} (\bibinfo {year} {2021})}\BibitemShut {NoStop}%
\bibitem [{\citenamefont {Chakraborty}\ \emph {et~al.}(2017)\citenamefont
  {Chakraborty}, \citenamefont {Tomsett}, \citenamefont {Raghavendra},
  \citenamefont {Harborne}, \citenamefont {Alzantot}, \citenamefont {Cerutti},
  \citenamefont {Srivastava}, \citenamefont {Preece}, \citenamefont {Julier},
  \citenamefont {Rao}, \citenamefont {Kelley}, \citenamefont {Braines},
  \citenamefont {Sensoy}, \citenamefont {Willis},\ and\ \citenamefont
  {Gurram}}]{chakraborty2017interpretability}%
  \BibitemOpen
  \bibfield  {author} {\bibinfo {author} {\bibfnamefont {S.}~\bibnamefont
  {Chakraborty}}, \bibinfo {author} {\bibfnamefont {R.}~\bibnamefont
  {Tomsett}}, \bibinfo {author} {\bibfnamefont {R.}~\bibnamefont
  {Raghavendra}}, \bibinfo {author} {\bibfnamefont {D.}~\bibnamefont
  {Harborne}}, \bibinfo {author} {\bibfnamefont {M.}~\bibnamefont {Alzantot}},
  \bibinfo {author} {\bibfnamefont {F.}~\bibnamefont {Cerutti}}, \bibinfo
  {author} {\bibfnamefont {M.}~\bibnamefont {Srivastava}}, \bibinfo {author}
  {\bibfnamefont {A.}~\bibnamefont {Preece}}, \bibinfo {author} {\bibfnamefont
  {S.}~\bibnamefont {Julier}}, \bibinfo {author} {\bibfnamefont {R.~M.}\
  \bibnamefont {Rao}}, \bibinfo {author} {\bibfnamefont {T.~D.}\ \bibnamefont
  {Kelley}}, \bibinfo {author} {\bibfnamefont {D.}~\bibnamefont {Braines}},
  \bibinfo {author} {\bibfnamefont {M.}~\bibnamefont {Sensoy}}, \bibinfo
  {author} {\bibfnamefont {C.~J.}\ \bibnamefont {Willis}}, \ and\ \bibinfo
  {author} {\bibfnamefont {P.}~\bibnamefont {Gurram}},\ }\bibfield  {title}
  {\enquote {\bibinfo {title} {Interpretability of deep learning models: A
  survey of results},}\ }\bibfield  {booktitle} {\emph {\bibinfo {booktitle}
  {2017 IEEE SmartWorld, Ubiquitous Intelligence \& Computing, Advanced \&
  Trusted Computed, Scalable Computing \& Communications, Cloud \& Big Data
  Computing, Internet of People and Smart City Innovation
  (SmartWorld/SCALCOM/UIC/ATC/CBDCom/IOP/SCI)}},\ }\href {\doibase
  10.1109/UIC-ATC.2017.8397411} {\bibfield  {journal} {\bibinfo  {journal}
  {IEEE SmartWorld}\ ,\ \bibinfo {pages} {1--6}} (\bibinfo {year}
  {2017})}\BibitemShut {NoStop}%
\bibitem [{\citenamefont {Zhang}\ and\ \citenamefont
  {Zhu}(2018)}]{zhang2018visual}%
  \BibitemOpen
  \bibfield  {author} {\bibinfo {author} {\bibfnamefont {Q.-S.}\ \bibnamefont
  {Zhang}}\ and\ \bibinfo {author} {\bibfnamefont {S.-C.}\ \bibnamefont
  {Zhu}},\ }\bibfield  {title} {\enquote {\bibinfo {title} {Visual
  interpretability for deep learning: a survey},}\ }\href@noop {} {\bibfield
  {journal} {\bibinfo  {journal} {Frontiers of Information Technology \&
  Electronic Engineering}\ }\textbf {\bibinfo {volume} {19}},\ \bibinfo {pages}
  {27--39} (\bibinfo {year} {2018})}\BibitemShut {NoStop}%
\bibitem [{\citenamefont {Shankaranarayana}\ and\ \citenamefont
  {Runje}(2019)}]{shankaranarayana2019alime}%
  \BibitemOpen
  \bibfield  {author} {\bibinfo {author} {\bibfnamefont {S.~M.}\ \bibnamefont
  {Shankaranarayana}}\ and\ \bibinfo {author} {\bibfnamefont {D.}~\bibnamefont
  {Runje}},\ }\bibfield  {title} {\enquote {\bibinfo {title} {{ALIME}:
  {A}utoencoder {B}ased {A}pproach for {L}ocal {I}nterpretability},}\
  }\bibfield  {booktitle} {\emph {\bibinfo {booktitle} {Intelligent Data
  Engineering and Automated Learning -- IDEAL 2019}},\ }\href@noop {} {\ ,\
  \bibinfo {pages} {454--463} (\bibinfo {year} {2019})}\BibitemShut {NoStop}%
\bibitem [{\citenamefont {Eckhardt}\ \emph {et~al.}(2007)\citenamefont
  {Eckhardt}, \citenamefont {Schneider}, \citenamefont {Hof},\ and\
  \citenamefont {Westerweel}}]{eckhardt2007turbulence}%
  \BibitemOpen
  \bibfield  {author} {\bibinfo {author} {\bibfnamefont {B.}~\bibnamefont
  {Eckhardt}}, \bibinfo {author} {\bibfnamefont {T.~M.}\ \bibnamefont
  {Schneider}}, \bibinfo {author} {\bibfnamefont {B.}~\bibnamefont {Hof}}, \
  and\ \bibinfo {author} {\bibfnamefont {J.}~\bibnamefont {Westerweel}},\
  }\bibfield  {title} {\enquote {\bibinfo {title} {Turbulence transition in
  pipe flow},}\ }\href@noop {} {\bibfield  {journal} {\bibinfo  {journal}
  {Annu. Rev. Fluid Mech.}\ }\textbf {\bibinfo {volume} {39}},\ \bibinfo
  {pages} {447--468} (\bibinfo {year} {2007})}\BibitemShut {NoStop}%
\bibitem [{\citenamefont {Fresca}\ and\ \citenamefont
  {Manzoni}(2022)}]{fresca2022pod}%
  \BibitemOpen
  \bibfield  {author} {\bibinfo {author} {\bibfnamefont {S.}~\bibnamefont
  {Fresca}}\ and\ \bibinfo {author} {\bibfnamefont {A.}~\bibnamefont
  {Manzoni}},\ }\bibfield  {title} {\enquote {\bibinfo {title}
  {{POD}-{DL}-{ROM}: {E}nhancing deep learning-based reduced order models for
  nonlinear parametrized {PDE}s by proper orthogonal decomposition},}\
  }\href@noop {} {\bibfield  {journal} {\bibinfo  {journal} {Computer Methods
  in Applied Mechanics and Engineering}\ }\textbf {\bibinfo {volume} {388}},\
  \bibinfo {pages} {114181} (\bibinfo {year} {2022})}\BibitemShut {NoStop}%
\bibitem [{\citenamefont {Murata}, \citenamefont {Fukami},\ and\ \citenamefont
  {Fukagata}(2020)}]{murata2020nonlinear}%
  \BibitemOpen
  \bibfield  {author} {\bibinfo {author} {\bibfnamefont {T.}~\bibnamefont
  {Murata}}, \bibinfo {author} {\bibfnamefont {K.}~\bibnamefont {Fukami}}, \
  and\ \bibinfo {author} {\bibfnamefont {K.}~\bibnamefont {Fukagata}},\
  }\bibfield  {title} {\enquote {\bibinfo {title} {Nonlinear mode decomposition
  with convolutional neural networks for fluid dynamics},}\ }\href@noop {}
  {\bibfield  {journal} {\bibinfo  {journal} {Journal of Fluid Mechanics}\
  }\textbf {\bibinfo {volume} {882}},\ \bibinfo {pages} {A13} (\bibinfo {year}
  {2020})}\BibitemShut {NoStop}%
\bibitem [{\citenamefont {Fukami}, \citenamefont {Nakamura},\ and\
  \citenamefont {Fukagata}(2020)}]{fukami2020convolutional}%
  \BibitemOpen
  \bibfield  {author} {\bibinfo {author} {\bibfnamefont {K.}~\bibnamefont
  {Fukami}}, \bibinfo {author} {\bibfnamefont {T.}~\bibnamefont {Nakamura}}, \
  and\ \bibinfo {author} {\bibfnamefont {K.}~\bibnamefont {Fukagata}},\
  }\bibfield  {title} {\enquote {\bibinfo {title} {{Convolutional neural
  network based hierarchical autoencoder for nonlinear mode decomposition of
  fluid field data}},}\ }\href {\doibase 10.1063/5.0020721} {\bibfield
  {journal} {\bibinfo  {journal} {Physics of Fluids}\ }\textbf {\bibinfo
  {volume} {32}},\ \bibinfo {pages} {095110} (\bibinfo {year} {2020})},\
  \Eprint
  {http://arxiv.org/abs/https://pubs.aip.org/aip/pof/article-pdf/doi/10.1063/5.0020721/14710996/095110\_1\_online.pdf}
  {https://pubs.aip.org/aip/pof/article-pdf/doi/10.1063/5.0020721/14710996/095110\_1\_online.pdf}
  \BibitemShut {NoStop}%
\bibitem [{\citenamefont {Chill{\`a}}\ and\ \citenamefont
  {Schumacher}(2012)}]{chilla2012new}%
  \BibitemOpen
  \bibfield  {author} {\bibinfo {author} {\bibfnamefont {F.}~\bibnamefont
  {Chill{\`a}}}\ and\ \bibinfo {author} {\bibfnamefont {J.}~\bibnamefont
  {Schumacher}},\ }\bibfield  {title} {\enquote {\bibinfo {title} {New
  perspectives in turbulent {R}ayleigh-{B}{\'e}nard convection},}\ }\href@noop
  {} {\bibfield  {journal} {\bibinfo  {journal} {The European Physical Journal
  E}\ }\textbf {\bibinfo {volume} {35}},\ \bibinfo {pages} {1--25} (\bibinfo
  {year} {2012})}\BibitemShut {NoStop}%
\bibitem [{\citenamefont {Verma}(2018)}]{verma2018physics}%
  \BibitemOpen
  \bibfield  {author} {\bibinfo {author} {\bibfnamefont {M.~K.}\ \bibnamefont
  {Verma}},\ }\href@noop {} {\emph {\bibinfo {title} {Physics of {B}uoyant
  {F}lows: {F}rom {I}nstabilities to {T}urbulence}}}\ (\bibinfo  {publisher}
  {World Scientific},\ \bibinfo {year} {2018})\BibitemShut {NoStop}%
\bibitem [{\citenamefont {Sharifi~Ghazijahani}\ \emph
  {et~al.}(2023)\citenamefont {Sharifi~Ghazijahani}, \citenamefont
  {K{\"a}stner}, \citenamefont {Valori}, \citenamefont {Thieme}, \citenamefont
  {T{\"a}schner}, \citenamefont {Schumacher},\ and\ \citenamefont
  {Cierpka}}]{sharifi2023scalex}%
  \BibitemOpen
  \bibfield  {author} {\bibinfo {author} {\bibfnamefont {M.}~\bibnamefont
  {Sharifi~Ghazijahani}}, \bibinfo {author} {\bibfnamefont {C.}~\bibnamefont
  {K{\"a}stner}}, \bibinfo {author} {\bibfnamefont {V.}~\bibnamefont {Valori}},
  \bibinfo {author} {\bibfnamefont {A.}~\bibnamefont {Thieme}}, \bibinfo
  {author} {\bibfnamefont {K.}~\bibnamefont {T{\"a}schner}}, \bibinfo {author}
  {\bibfnamefont {J.}~\bibnamefont {Schumacher}}, \ and\ \bibinfo {author}
  {\bibfnamefont {C.}~\bibnamefont {Cierpka}},\ }\bibfield  {title} {\enquote
  {\bibinfo {title} {The {SCALEX} facility--an apparatus for scaled fluid
  dynamical experiments},}\ }\href@noop {} {\bibfield  {journal} {\bibinfo
  {journal} {tm-Technisches Messen}\ }\textbf {\bibinfo {volume} {90}},\
  \bibinfo {pages} {296--309} (\bibinfo {year} {2023})}\BibitemShut {NoStop}%
\bibitem [{\citenamefont {Sharifi~Ghazijahani}\ and\ \citenamefont
  {Cierpka}(2024)}]{sharifi2024spatio}%
  \BibitemOpen
  \bibfield  {author} {\bibinfo {author} {\bibfnamefont {M.}~\bibnamefont
  {Sharifi~Ghazijahani}}\ and\ \bibinfo {author} {\bibfnamefont
  {C.}~\bibnamefont {Cierpka}},\ }\bibfield  {title} {\enquote {\bibinfo
  {title} {{Spatio-temporal dynamics of superstructures and vortices in
  turbulent {R}ayleigh–{B}{é}nard convection}},}\ }\href {\doibase
  10.1063/5.0191403} {\bibfield  {journal} {\bibinfo  {journal} {Physics of
  Fluids}\ }\textbf {\bibinfo {volume} {36}},\ \bibinfo {pages} {035120}
  (\bibinfo {year} {2024})},\ \Eprint
  {http://arxiv.org/abs/https://pubs.aip.org/aip/pof/article-pdf/doi/10.1063/5.0191403/19716786/035120\_1\_5.0191403.pdf}
  {https://pubs.aip.org/aip/pof/article-pdf/doi/10.1063/5.0191403/19716786/035120\_1\_5.0191403.pdf}
  \BibitemShut {NoStop}%
\bibitem [{\citenamefont {Perez}\ and\ \citenamefont
  {Wang}(2017)}]{perez2017effectiveness}%
  \BibitemOpen
  \bibfield  {author} {\bibinfo {author} {\bibfnamefont {L.}~\bibnamefont
  {Perez}}\ and\ \bibinfo {author} {\bibfnamefont {J.}~\bibnamefont {Wang}},\
  }\bibfield  {title} {\enquote {\bibinfo {title} {The effectiveness of data
  augmentation in image classification using deep learning},}\ }\href@noop {}
  {\bibfield  {journal} {\bibinfo  {journal} {arXiv preprint arXiv:1712.04621}\
  } (\bibinfo {year} {2017})}\BibitemShut {NoStop}%
\bibitem [{\citenamefont {Cubuk}\ \emph {et~al.}(2019)\citenamefont {Cubuk},
  \citenamefont {Zoph}, \citenamefont {Mané}, \citenamefont {Vasudevan},\ and\
  \citenamefont {Le}}]{cubuk2019autoaugment}%
  \BibitemOpen
  \bibfield  {author} {\bibinfo {author} {\bibfnamefont {E.~D.}\ \bibnamefont
  {Cubuk}}, \bibinfo {author} {\bibfnamefont {B.}~\bibnamefont {Zoph}},
  \bibinfo {author} {\bibfnamefont {D.}~\bibnamefont {Mané}}, \bibinfo
  {author} {\bibfnamefont {V.}~\bibnamefont {Vasudevan}}, \ and\ \bibinfo
  {author} {\bibfnamefont {Q.~V.}\ \bibnamefont {Le}},\ }\bibfield  {title}
  {\enquote {\bibinfo {title} {Auto{A}ugment: {L}earning augmentation
  strategies from data},}\ }\bibfield  {booktitle} {\emph {\bibinfo {booktitle}
  {2019 IEEE/CVF Conference on Computer Vision and Pattern Recognition
  (CVPR)}},\ }\href {\doibase 10.1109/CVPR.2019.00020} {\bibfield  {journal}
  {\bibinfo  {journal} {CVPR}\ ,\ \bibinfo {pages} {113--123}} (\bibinfo {year}
  {2019})}\BibitemShut {NoStop}%
\bibitem [{\citenamefont {Yang}\ \emph {et~al.}(2022)\citenamefont {Yang},
  \citenamefont {Xiao}, \citenamefont {Zhang}, \citenamefont {Guo},
  \citenamefont {Zhao},\ and\ \citenamefont {Shen}}]{yang2022image}%
  \BibitemOpen
  \bibfield  {author} {\bibinfo {author} {\bibfnamefont {S.}~\bibnamefont
  {Yang}}, \bibinfo {author} {\bibfnamefont {W.}~\bibnamefont {Xiao}}, \bibinfo
  {author} {\bibfnamefont {M.}~\bibnamefont {Zhang}}, \bibinfo {author}
  {\bibfnamefont {S.}~\bibnamefont {Guo}}, \bibinfo {author} {\bibfnamefont
  {J.}~\bibnamefont {Zhao}}, \ and\ \bibinfo {author} {\bibfnamefont
  {F.}~\bibnamefont {Shen}},\ }\bibfield  {title} {\enquote {\bibinfo {title}
  {Image data augmentation for deep learning: A survey},}\ }\href@noop {}
  {\bibfield  {journal} {\bibinfo  {journal} {arXiv preprint arXiv:2204.08610}\
  } (\bibinfo {year} {2022})}\BibitemShut {NoStop}%
\bibitem [{\citenamefont {Teutsch}\ \emph {et~al.}(2024)\citenamefont
  {Teutsch}, \citenamefont {Ghazijahani}, \citenamefont {Heyder}, \citenamefont
  {Cierpka}, \citenamefont {Schumacher},\ and\ \citenamefont
  {M{\"a}der}}]{teutsch2024large}%
  \BibitemOpen
  \bibfield  {author} {\bibinfo {author} {\bibfnamefont {P.}~\bibnamefont
  {Teutsch}}, \bibinfo {author} {\bibfnamefont {M.~S.}\ \bibnamefont
  {Ghazijahani}}, \bibinfo {author} {\bibfnamefont {F.}~\bibnamefont {Heyder}},
  \bibinfo {author} {\bibfnamefont {C.}~\bibnamefont {Cierpka}}, \bibinfo
  {author} {\bibfnamefont {J.}~\bibnamefont {Schumacher}}, \ and\ \bibinfo
  {author} {\bibfnamefont {P.}~\bibnamefont {M{\"a}der}},\ }\bibfield  {title}
  {\enquote {\bibinfo {title} {Large-scale-aware data augmentation for
  reduced-order models of high-dimensional flows},}\ }\href@noop {} {\
  (\bibinfo {year} {2024})},\ \bibinfo {note} {{M}anuscript under
  review}\BibitemShut {NoStop}%
\bibitem [{\citenamefont {Liang}\ \emph {et~al.}(2002)\citenamefont {Liang},
  \citenamefont {Lee}, \citenamefont {Lim}, \citenamefont {Lin}, \citenamefont
  {Lee},\ and\ \citenamefont {Wu}}]{LIANG_POD}%
  \BibitemOpen
  \bibfield  {author} {\bibinfo {author} {\bibfnamefont {Y.}~\bibnamefont
  {Liang}}, \bibinfo {author} {\bibfnamefont {H.}~\bibnamefont {Lee}}, \bibinfo
  {author} {\bibfnamefont {S.}~\bibnamefont {Lim}}, \bibinfo {author}
  {\bibfnamefont {W.}~\bibnamefont {Lin}}, \bibinfo {author} {\bibfnamefont
  {K.}~\bibnamefont {Lee}}, \ and\ \bibinfo {author} {\bibfnamefont
  {C.}~\bibnamefont {Wu}},\ }\bibfield  {title} {\enquote {\bibinfo {title}
  {Proper {O}rthogonal {D}ecomposition and its {Applications} — {P}art {I}:
  {T}heory},}\ }\href {\doibase https://doi.org/10.1006/jsvi.2001.4041}
  {\bibfield  {journal} {\bibinfo  {journal} {Journal of Sound and Vibration}\
  }\textbf {\bibinfo {volume} {252}},\ \bibinfo {pages} {527--544} (\bibinfo
  {year} {2002})}\BibitemShut {NoStop}%
\bibitem [{\citenamefont {Maas}\ \emph {et~al.}(2013)\citenamefont {Maas},
  \citenamefont {Hannun}, \citenamefont {Ng} \emph
  {et~al.}}]{maas2013rectifier}%
  \BibitemOpen
  \bibfield  {author} {\bibinfo {author} {\bibfnamefont {A.~L.}\ \bibnamefont
  {Maas}}, \bibinfo {author} {\bibfnamefont {A.~Y.}\ \bibnamefont {Hannun}},
  \bibinfo {author} {\bibfnamefont {A.~Y.}\ \bibnamefont {Ng}},  \emph
  {et~al.},\ }\bibfield  {title} {\enquote {\bibinfo {title} {Rectifier
  nonlinearities improve neural network acoustic models},}\ }\bibfield
  {booktitle} {\emph {\bibinfo {booktitle} {Proc. ICML}},\ }\href@noop {} {\
  \textbf {\bibinfo {volume} {30}},\ \bibinfo {pages} {3} (\bibinfo {year}
  {2013})}\BibitemShut {NoStop}%
\bibitem [{\citenamefont {Ioffe}\ and\ \citenamefont
  {Szegedy}(2015)}]{ioffe2015batch}%
  \BibitemOpen
  \bibfield  {author} {\bibinfo {author} {\bibfnamefont {S.}~\bibnamefont
  {Ioffe}}\ and\ \bibinfo {author} {\bibfnamefont {C.}~\bibnamefont
  {Szegedy}},\ }\bibfield  {title} {\enquote {\bibinfo {title} {Batch
  {N}ormalization: {A}ccelerating {D}eep {N}etwork {T}raining by {R}educing
  {I}nternal {C}ovariate {S}hift},}\ }\bibfield  {booktitle} {\emph {\bibinfo
  {booktitle} {Proceedings of the 32nd International Conference on Machine
  Learning}},\ }\href {https://proceedings.mlr.press/v37/ioffe15.html}
  {\bibfield  {journal} {\bibinfo  {journal} {Proceedings of Machine Learning
  Research}\ }\textbf {\bibinfo {volume} {37}},\ \bibinfo {pages} {448--456}
  (\bibinfo {year} {2015})}\BibitemShut {NoStop}%
\bibitem [{\citenamefont {Srivastava}\ \emph {et~al.}(2014)\citenamefont
  {Srivastava}, \citenamefont {Hinton}, \citenamefont {Krizhevsky},
  \citenamefont {Sutskever},\ and\ \citenamefont
  {Salakhutdinov}}]{srivastava2014dropout}%
  \BibitemOpen
  \bibfield  {author} {\bibinfo {author} {\bibfnamefont {N.}~\bibnamefont
  {Srivastava}}, \bibinfo {author} {\bibfnamefont {G.}~\bibnamefont {Hinton}},
  \bibinfo {author} {\bibfnamefont {A.}~\bibnamefont {Krizhevsky}}, \bibinfo
  {author} {\bibfnamefont {I.}~\bibnamefont {Sutskever}}, \ and\ \bibinfo
  {author} {\bibfnamefont {R.}~\bibnamefont {Salakhutdinov}},\ }\bibfield
  {title} {\enquote {\bibinfo {title} {Dropout: {A} simple way to prevent
  neural networks from overfitting},}\ }\href@noop {} {\bibfield  {journal}
  {\bibinfo  {journal} {The Journal of Machine Learning Research}\ }\textbf
  {\bibinfo {volume} {15}},\ \bibinfo {pages} {1929--1958} (\bibinfo {year}
  {2014})}\BibitemShut {NoStop}%
\bibitem [{\citenamefont {Bachman}, \citenamefont {Alsharif},\ and\
  \citenamefont {Precup}(2014)}]{bachman2014learning}%
  \BibitemOpen
  \bibfield  {author} {\bibinfo {author} {\bibfnamefont {P.}~\bibnamefont
  {Bachman}}, \bibinfo {author} {\bibfnamefont {O.}~\bibnamefont {Alsharif}}, \
  and\ \bibinfo {author} {\bibfnamefont {D.}~\bibnamefont {Precup}},\
  }\bibfield  {title} {\enquote {\bibinfo {title} {Learning with
  {P}seudo-{E}nsembles},}\ }\href@noop {} {\bibfield  {journal} {\bibinfo
  {journal} {Advances in Neural Information Processing Systems}\ }\textbf
  {\bibinfo {volume} {27}} (\bibinfo {year} {2014})}\BibitemShut {NoStop}%
\bibitem [{\citenamefont {Hara}, \citenamefont {Saitoh},\ and\ \citenamefont
  {Shouno}(2016)}]{hara2016analysis}%
  \BibitemOpen
  \bibfield  {author} {\bibinfo {author} {\bibfnamefont {K.}~\bibnamefont
  {Hara}}, \bibinfo {author} {\bibfnamefont {D.}~\bibnamefont {Saitoh}}, \ and\
  \bibinfo {author} {\bibfnamefont {H.}~\bibnamefont {Shouno}},\ }\bibfield
  {title} {\enquote {\bibinfo {title} {Analysis of dropout learning regarded as
  ensemble learning},}\ }\bibfield  {booktitle} {\emph {\bibinfo {booktitle}
  {Artificial Neural Networks and Machine Learning -- ICANN 2016}},\
  }\href@noop {} {\bibfield  {journal} {\bibinfo  {journal} {ICANN}\ ,\
  \bibinfo {pages} {72--79}} (\bibinfo {year} {2016})}\BibitemShut {NoStop}%
\bibitem [{\citenamefont {Kingma}\ and\ \citenamefont
  {Ba}(2014)}]{kingma2014adam}%
  \BibitemOpen
  \bibfield  {author} {\bibinfo {author} {\bibfnamefont {D.~P.}\ \bibnamefont
  {Kingma}}\ and\ \bibinfo {author} {\bibfnamefont {J.}~\bibnamefont {Ba}},\
  }\bibfield  {title} {\enquote {\bibinfo {title} {Adam: {A} method for
  stochastic optimization},}\ }\href@noop {} {\bibfield  {journal} {\bibinfo
  {journal} {arXiv preprint arXiv:1412.6980}\ } (\bibinfo {year}
  {2014})}\BibitemShut {NoStop}%
\bibitem [{\citenamefont {Bengio}\ \emph {et~al.}(2009)\citenamefont {Bengio},
  \citenamefont {Louradour}, \citenamefont {Collobert},\ and\ \citenamefont
  {Weston}}]{bengio2009curriculum}%
  \BibitemOpen
  \bibfield  {author} {\bibinfo {author} {\bibfnamefont {Y.}~\bibnamefont
  {Bengio}}, \bibinfo {author} {\bibfnamefont {J.}~\bibnamefont {Louradour}},
  \bibinfo {author} {\bibfnamefont {R.}~\bibnamefont {Collobert}}, \ and\
  \bibinfo {author} {\bibfnamefont {J.}~\bibnamefont {Weston}},\ }\bibfield
  {title} {\enquote {\bibinfo {title} {Curriculum learning},}\ }\bibfield
  {booktitle} {\emph {\bibinfo {booktitle} {Proc. ICML}},\ }\href {\doibase
  10.1145/1553374.1553380} {\ \bibinfo {series} {ICML '09},\ \bibinfo {pages}
  {41–48} (\bibinfo {year} {2009})}\BibitemShut {NoStop}%
\bibitem [{\citenamefont {Teutsch}\ and\ \citenamefont
  {M{\"a}der}(2022)}]{teutsch2022flipped}%
  \BibitemOpen
  \bibfield  {author} {\bibinfo {author} {\bibfnamefont {P.}~\bibnamefont
  {Teutsch}}\ and\ \bibinfo {author} {\bibfnamefont {P.}~\bibnamefont
  {M{\"a}der}},\ }\bibfield  {title} {\enquote {\bibinfo {title} {Flipped
  classroom: Effective teaching for time series forecasting},}\ }\href
  {https://openreview.net/forum?id=w3x20YEcQK} {\bibfield  {journal} {\bibinfo
  {journal} {Transactions on Machine Learning Research}\ } (\bibinfo {year}
  {2022})}\BibitemShut {NoStop}%
\end{thebibliography}%

\end{document}